\documentclass {jaa}
\usepackage[english]{babel}
\usepackage{times,amssymb,subfigure,relsize}
\usepackage{color,rotating}
\usepackage{amssymb}
\usepackage{comment}
\usepackage{graphicx}
\usepackage{hyperref}
\usepackage{graphics}
\usepackage{graphicx}
\usepackage{color}
\usepackage{amsmath}
\usepackage{psfrag}
\usepackage{epsfig}
\usepackage{multirow}
\usepackage{authblk}

\usepackage{placeins}
\begin{document}
\title{Large Area X-ray Proportional Counter (LAXPC) in Orbit Performance : Calibration, background, analysis software}
\author{H. M. Antia\textsuperscript{1,*}, P. C. Agrawal\textsuperscript{2},
	Dhiraj Dedhia\textsuperscript{1},
	Tilak Katoch\textsuperscript{1},
	R. K. Manchanda\textsuperscript{3},
Ranjeev Misra\textsuperscript{4},
	Kallol Mukerjee\textsuperscript{1},
Mayukh Pahari\textsuperscript{5,6},
Jayashree Roy\textsuperscript{4},
	P. Shah\textsuperscript{1},
	J. S. Yadav\textsuperscript{7}}
\affilOne{\textsuperscript{1}Department of Astronomy and Astrophysics, Tata Institute of Fundamental Research, Homi Bhabha Road, Mumbai 400005, India\\}
\affilTwo{\textsuperscript{2}Department of Astronomy and Astrophysics (retd), Tata Institute of Fundamental Research, Homi Bhabha Road, Mumbai 400005, India\\}
\affilThree{\textsuperscript{3}Centre for Astrophysics, University of Southern Queensland, QLD 4300, Australia\\}
\affilFour{\textsuperscript{4}Inter-University Centre for Astronomy \& Astrophysics, Ganeshkhind, Pune-411007, India\\}
\affilFive{\textsuperscript{5}School of Physics and Astronomy, University of Southampton, Highfield Campus, Southampton SO17 1BJ, UK\\}
\affilSix{\textsuperscript{6}Department of Physics, Indian Institute of Technology, Hyderabad 502285, India\\}
\affilSeven{\textsuperscript{7}Department of Physics, Indian Institute of Technology, Kanpur 208016, India\\}
\corres{antia@tifr.res.in}

\twocolumn[{

\maketitle



\corres{antia@tifr.res.in}

\begin{abstract}
The Large Area X-ray Proportional Counter (LAXPC) instrument on-board AstroSat has
three nominally identical detectors for timing and spectral studies in the energy range of 3--80 keV.
The performance of these detectors during the five years after the launch of AstroSat
is described. Currently, only one of the detector is working nominally.
The variation in pressure, energy resolution, gain and background with time are
discussed. The capabilities and limitations of the instrument are described.
A brief account of available analysis software is also provided.

\end{abstract}


}]

\section{Introduction}    

The Large Area X-ray Proportional Counter (LAXPC) instrument on-board AstroSat (Agrawal 2006; Singh et al.~2014) consists of
three co-aligned detectors for X-ray timing and spectral studies over an energy range of
3--80 keV (Yadav et al.~2016a; Agrawal et al.~2017). 
Apart from LAXPC, AstroSat has three more
co-aligned instruments, the Soft X-ray Telescope (SXT, Singh et al. 2016), the Cadmium Zinc Telluride
Imager (CZTI, Bhalerao et al.~2017) and the Ultra-Violet Imaging Telescope (UVIT,
Tandon et al.~2017).
AstroSat was conceived to carry out multiwavelength observations of various sources in
the Visible, UV and X-ray bands using these co-aligned instruments.
AstroSat was launched on September 28, 2015 and the initial calibration of the LAXPC instrument was discussed by Antia et al.~(2017). By now AstroSat has made more than 2000
distinct observations covering a wide variety of sources and a large amount of data are
publicly available from the AstroSat Data Archive\footnote{https://astrobrowse.issdc.gov.in/astro\_archive/archive/Home.jsp}.
A quick look light-curves of all LAXPC observations are available at the LAXPC website\footnote{https::www.tifr.res.in/\~{}astrosat\_laxpc/laxpclog.lc-hdr.html}.
A number of science results from LAXPC instrument have been published and a summary of
these results is described in the companion paper (Yadav et al.~2020).

Each LAXPC detector has five layers divided into seven anodes (A1--A7), with the two top layers having two
anodes each. In addition there are three veto anodes (A8--A10) on the three sides of the detector.
The three faces covered by veto anodes are, the bottom and the two faces covering the long side
of the detector as shown in Figure 2 of Antia et al.~(2017).
By default the LAXPC detectors operate in the Event Analysis (EA) mode where the timing and
energy of each photon is recorded with the time-resolution of 10 $\mu$s. The EA mode operation
also generates
the Broad Band Counting (BB) mode data, which gives the
counts with a predefined time-bin, for various energy bins and anodes, including the counts of events beyond the
Upper Level Discriminator (ULD) threshold (nominally at 80 keV). The dead-time of the detectors is 42 $\mu$s
(Yadav et al.~2016b).
In addition, there is a Fast Counter (FC) mode with a dead-time of about 10~$\mu$s to allow for
observation of bright sources. In this mode only the counts with a fixed time-bin of 160 $\mu$s,
as recorded in only the top layer of the detector in four predefined energy bins
are available. However, this mode has not been used for any science observation and
no software to analyse the data is available. As a result, this mode is not allowed to be
configured by LAXPC users. Thus effectively only one mode covering both EA and BB is allowed.

In the xenon gas counters, a large fraction of incoming photons above 34.5 keV emit a fluorescence photon
and depending on the cell geometry and filling pressure, the fluorescent photon may generate a
second localized electron cloud in a different anode. Therefore, on board electronics is
designed to recognise such correlated double events and the energy deposited in the two
anodes is added. These are referred to as double events as opposed to single events,
where all energy is deposited in one anode.

In this article, we mainly focus on the performance of the LAXPC instrument in orbit and
its calibration and some software which is available for analysing data. The three
LAXPC detectors are labelled as LAXPC10, LAXPC20 and LAXPC30. Currently, only
LAXPC20 is working nominally.  The rest of the paper is organised as follows:
Section 2 describes the performance of detectors during the last five years. Section 3
describes the variation in background with time and some procedures to correct for these.
Section 4 describes some capabilities and limitations of the detectors and their sensitivity.
Section 5 describes some
software for analysing the data. Section 6 describes the various science goals of
LAXPC detectors and how they are met by the data obtained so far.
Finally, Section 7 describes a summary of calibration and performance of detectors.

\begin{figure*} [th]
\centering
\includegraphics[width=0.95\columnwidth,angle=0.0]{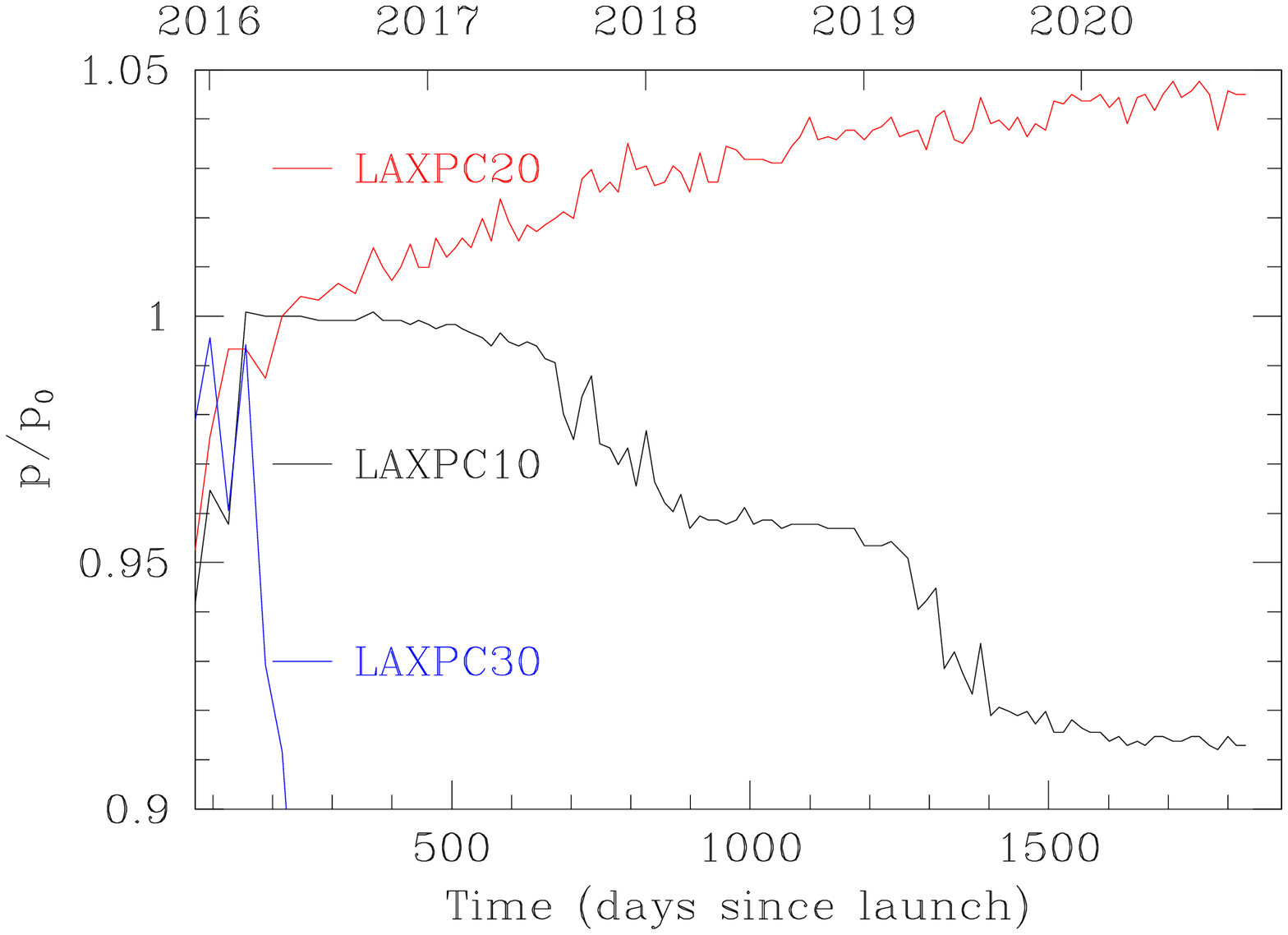}\quad
\includegraphics[width=0.95\columnwidth,angle=0.0]{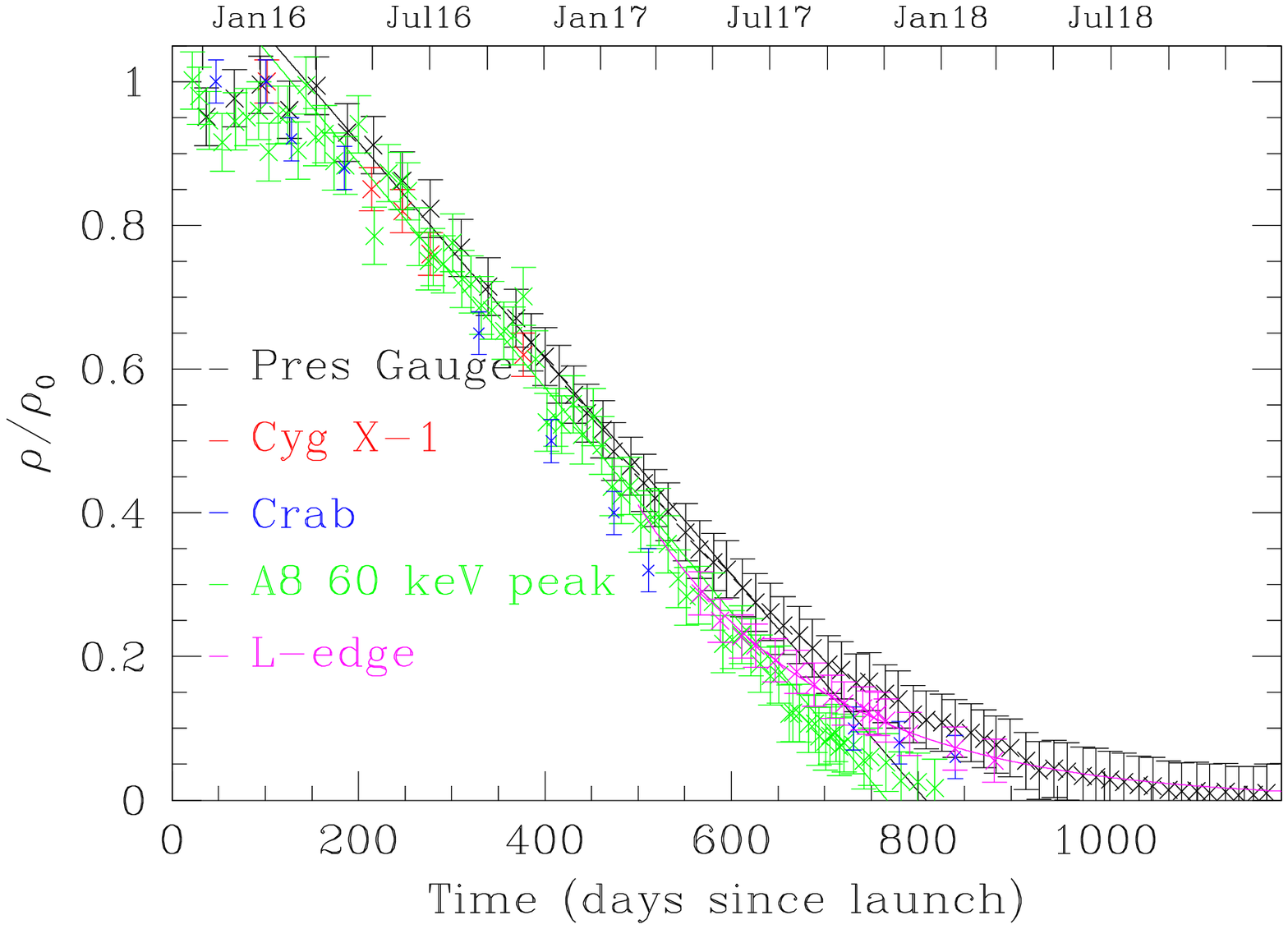}
\caption{The pressure in LAXPC detectors as a function of time. The left panel shows
the pressure as obtained from the pressure gauge for all three detectors.
The right panel shows the density for LAXPC30 using various techniques. The dashed-line
is the fit to linear part of the curve. Both pressure and density are shown relative to
their initial values.}
\label{fig:pres}
\end{figure*}

\section{Long Term Performance of LAXPC in Orbit}

The health parameters of the detectors, like the temperature, pressure, high voltage (HV)
and various energy thresholds are monitored regularly. While the temperature of detectors is
steady, the pressure has changed over the last five years from the initial value of
about two atmospheres. The HV and thresholds have been maintained, except for some adjustments
that were made from time to time. Further, to monitor the stability of detectors, the
peak position and energy resolution of 30 and 60 keV peaks in the veto anode A8 from the
on board radioactive source Am$^{241}$, are measured regularly and a log is maintained\footnote{https::www.tifr.res.in/\~{}astrosat\_laxpc/LaxpcSoft\_v1.0/gaina8.dat}.
Apart from these, we have also regularly monitored spectrum from Crab observation to check the stability
of the detector response.

\subsection{Pressure in Detectors}

Figure~\ref{fig:pres}, shows the pressure as estimated from the pressure gauge in all detectors as a function of time.
The LAXPC30 developed a leak
soon after launch and the pressure was decreasing steadily. The HV of the
detector was turned off on March 8, 2018 when the pressure had reduced to about 5\% of the
original pressure. LAXPC10 also has a fine leak and the pressure has
been reducing gradually.
Curiously, the pressure gauge of LAXPC20 shows a slow increase
in pressure. This is most likely an artefact and shows the limitation of on-board pressure
gauge. Because of this, we used other techniques to estimate the density in
LAXPC30, as described by Antia et al.~(2017) and the results using these are shown in
the right panel of Figure~\ref{fig:pres}. Since these techniques are based on absorption
in Xenon gas, they yield the density which is assumed to be a proxy for pressure, as
the temperature is almost constant during the entire period. The Cyg X-1 observations
during the soft state were used to measure the density by calibrating the ratio of
counts around 20 keV, observed in different layers of the detector. The soft state was used
to ensure very low flux beyond the Xe K-edge to avoid possible contamination from
events involving Xe fluorescence photons. The Crab spectra were fitted using
responses with different density to get the best fit for density. Other techniques were
based on the strength of 60 keV peak in veto anode A8 and the observation of L-edge in
the spectrum when the density was sufficiently low.
It can be seen that results from all these techniques agree with each other.
The density in LAXPC30 decreased linearly for
some time at a rate of about 4.5\% of original density per month. After that it
followed an exponential rate, as may be expected for a leak, with an e-folding time of
about 200 days. By now the pressure is below the sensitivity of pressure gauge and
using the exponential profile it can be estimated to be about 0.05\% of the original value.
To maintain the gain of the detector in a reasonable limit the HV of detector was
reduced from time to time.
To allow analysis of LAXPC30 data, the response has been calculated at different densities
and the software gives recommendation of which response to use depending on the
time of observation.

\subsection{Energy Resolution and Gain of Detectors}

\begin{figure*} [th]
\centering
\includegraphics[width=0.95\columnwidth,angle=0.0]{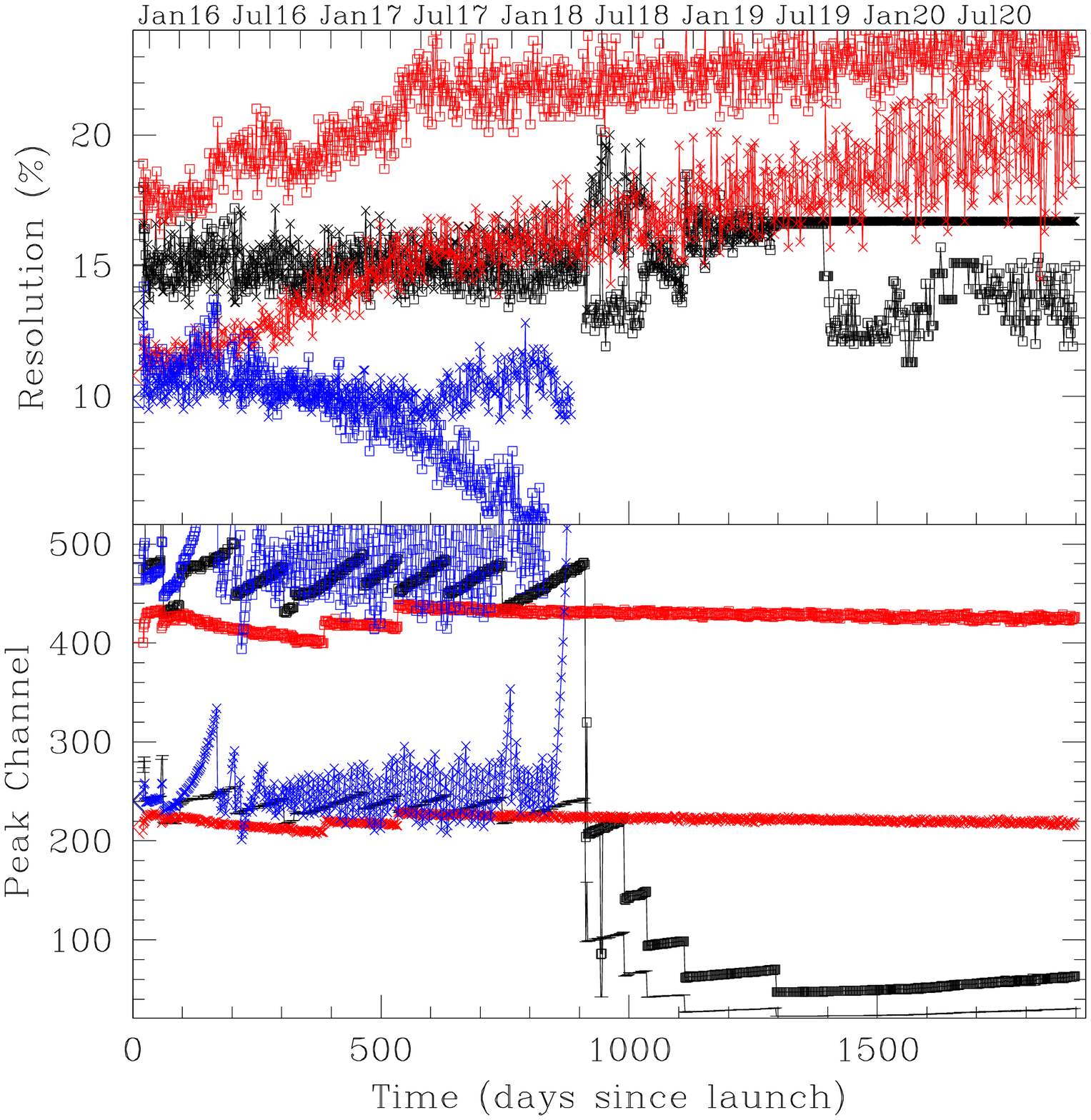}\quad
\includegraphics[width=0.95\columnwidth,angle=0.0]{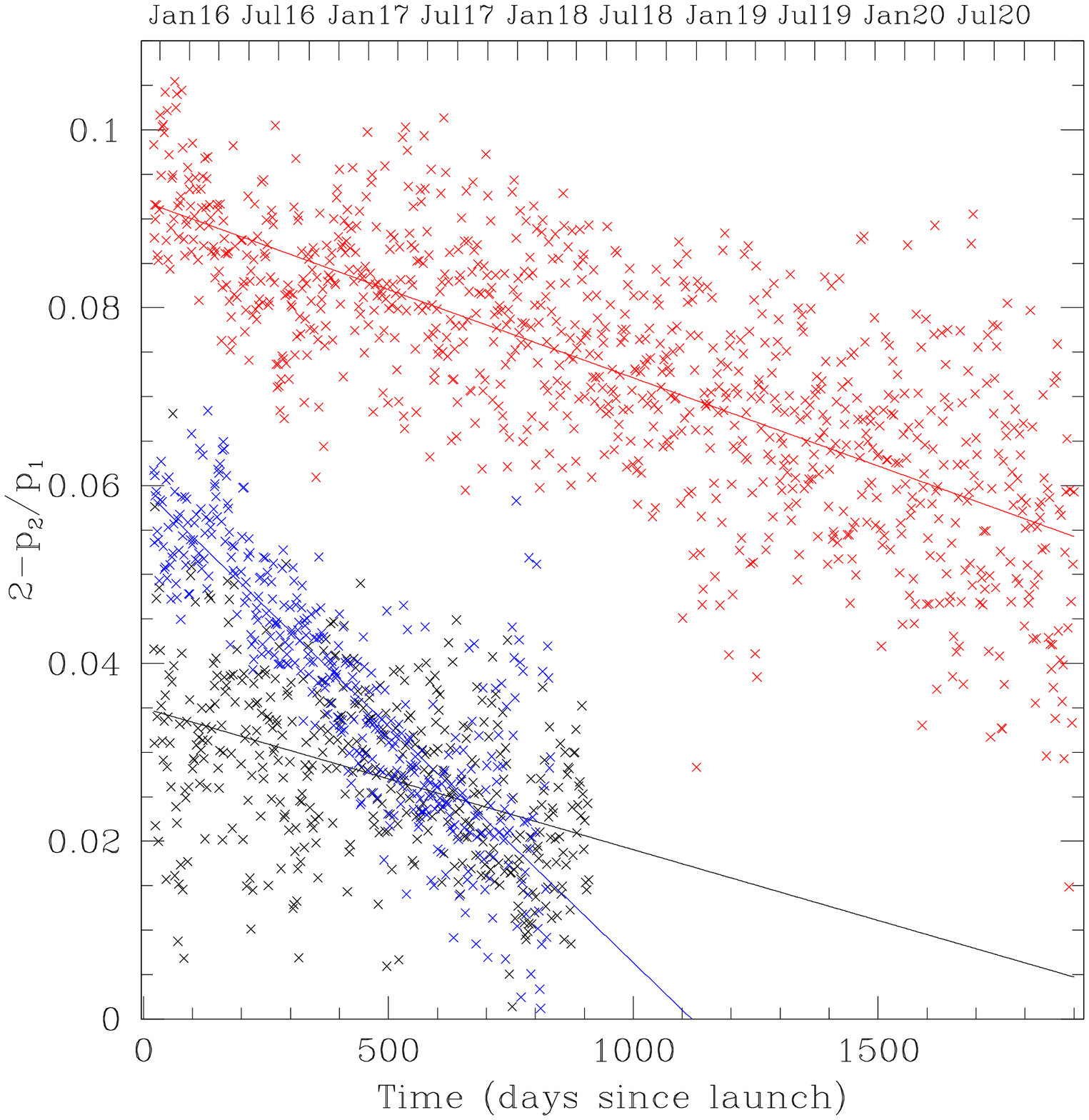}
\caption{The energy resolution and peak channel for the 30 keV ($p_1$, as marked by cross) and 60 keV
($p_2$, as marked by open squares) calibration
peak in veto anode A8 is shown in the left panel. The black, red and blue points show
the result for LAXPC10, LAXPC20 and LAXPC30, respectively. The right panel shows
the quantity $2-p_2/p_1$ for the three detectors. The lines mark the best fit straight
lines for the three detectors.}
\label{fig:a8}
\end{figure*}

Because of the leak, the gain of LAXPC10 and LAXPC30 was also increasing with time and this
needs to be estimated using the 30 keV line in the veto anode A8 from the Am$^{241}$
source. The calibration source has two lines, one at 30 keV due to K-escape event from Xe and
another at 60 keV from the Am$^{241}$ source, and these peaks can
be used to check the drift in gain as well as energy resolution. To correct for the drift in the gain, the HV of
LAXPC10 and LAXPC30 was adjusted from time to time. This gives some steps in the gain.
After some stage the gain of LAXPC30 had to be adjusted frequently, giving a band in the peak
position as seen in the left panel of Figure~\ref{fig:a8}. On January 22, 2018 the HV of LAXPC30 was reduced
to the minimum possible value of about 930 V (as compared to the initial value of about 2300 V), after which the peak channel kept shifting upwards.
Even before this stage, the 60 keV peak was not well defined due to low efficiency and hence its position
could not be determined.
By the time the HV of detector was turned off, even the 30 keV peak had shifted beyond the
ULD and it was not possible to estimate the gain
of the detector.

The channel to energy mapping is defined by
a quad\-ratic (Antia et al.~2017) and it is not possible to estimate the three coefficients of
quadratic using the peak position of two peaks. Hence, it is assumed that only the linear
term is changing with time and its value is estimated by the position of the 30 keV peak.
Responses for different values of the peak position of 30 keV peak are provided and software
makes appropriate recommendation based on the time of observation. If the gain was linear
the peak channel for 60 keV peak, $p_2$ would be twice that for 30 keV peak, $p_1$. Hence,
the difference $2-p_2/p_1$ gives a measure of non-linearity in the gain. This quantity is
shown in the right panel of Figure~\ref{fig:a8}. It can be seen that the nonlinearity
has been decreasing for all detectors. However, it should be noted that this quantity can
also change if the constant term in the gain is changing. Thus it is not possible to
correct for this variation. It is advisable to use {\tt gain fit} command in {\tt Xspec} to adjust the
constant term, and even the linear term, to get the best spectral fit.

On March  26, 2018, LAXPC10 showed erratic counts with strong bursts where the dead-time corrected
count rate reached 40000 s$^{-1}$. The cause of this anomaly is not known.
To stabilise the counts, the HV of the detector was reduced.
Attempts were also made to control
the noise by adjusting the Low Level Discrimination (LLD) thresholds of some anodes which were showing low channel noise,
but that did not remove the bursts and hence the HV was kept at lower value.
After that the counts were stable to some extent though smaller
bursts continued. By looking at the value of counts beyond the ULD
it is possible to identify
the time intervals when counts are not stable and this has been implemented in the software
which automatically removes these time intervals from Good Time Intervals (GTI). This problem occurred a few
times after that and every time the HV was reduced to stabilise the counts. The last
adjustment was made on April 9, 2019 and since then no bursts have been observed,
except for a few days between April 23 and May 3, 2020. Because of these adjustments, the HV
of LAXPC10 has been reduced from about 2330 V initially to about 2190 V in March 2018
(before the anomalous behaviour started) and to 1860 V now. Some reduction in HV
was also required to compensate for the leak in the detector.
About half of the 470~V
reduction would have been needed to compensate for the reduction in pressure.
The status of LAXPC10 at 
any time can be checked on the LAXPC website\footnote{https::www.tifr.res.in/\~{}astrosat\_laxpc/laxpc10.pdf}.
Even if the detector remains stable, it would take a few years for it to reach a reasonable gain
due to leakage.

Because of the reduction in HV the energy thresholds
of LAXPC10 have increased. At the time of last adjustment, the LLD was around 30 keV
and ULD was about 400 keV. Since then because of the fine leak the thresholds have reduced to
some extent, with LLD around 22 keV and ULD around 320 keV. It is difficult to estimate
the gain of this detector reliably and to get the corresponding response. As a result,
it is not possible to use this detector for spectroscopic studies. During the period
just after March 26, 2018 the gain of the detector was in a reasonable range.
As a result, the data obtained during that interval could be analysed if
single event mode for only the top layer of the detector is used.
It is necessary to reject all double events where the energy is deposited in two different
anodes due to Xe K X-ray photons being absorbed in a different anode, as
the energy thresholds for choosing these events have not been adjusted due to
difficulty in estimating them reliably. The restriction to top layer of detector is needed
because the LLD threshold of some other anodes has been increased giving an edge in the
spectrum.

 \begin{figure} [thb]
\centering
\includegraphics[width=0.95\columnwidth,angle=0.0]{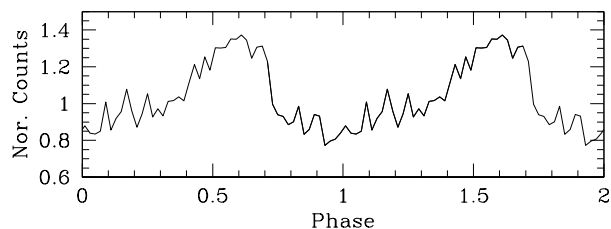}\quad
\caption{The pulse profile of GRO J2058+42 as calculated from LAXPC10 data.}
\label{fig:pulse}
\end{figure}

Even with the low gain, LAXPC10 does detect bursts, e.g., from GRB (Antia et al.~2020a,b).
Similarly, it is possible to detect pulsation in LAXPC10, e.g., for GRO J2058+42 during
its outburst in April 2019, the pulsation period was determined to be $194.256\pm 0.034$ s
and spin-up rate was estimate to be $\dot\nu=(1.7\pm 1.0)\times10^{-11}$ Hz s$^{-1}$. This can
be compared with $P=194.2201\pm 0.0016$ s and $\dot\nu=(1.65\pm0.06)\times10^{-11}$ Hz
s$^{-1}$ obtained with LAXPC20 (Mukerjee et al.~2020a). The error-bars represent the 90\%
confidence limits. The higher error in LAXPC10 is mainly due
to lower counts because of higher LLD and low efficiency. This observation was taken
at a time when the counts were not very stable in LAXPC10 and only
about 7\% of exposure time was usable. The pulse profile obtained from LAXPC10 is shown in
Figure~\ref{fig:pulse}. The gain of the detector during this observation is uncertain
and the LLD was probably around 30 keV. This pulse profile can be compared with pulse profile obtained for energy range
30--40 keV from LAXPC20 (Mukerjee et al.~2020a).

 \begin{figure} [H]
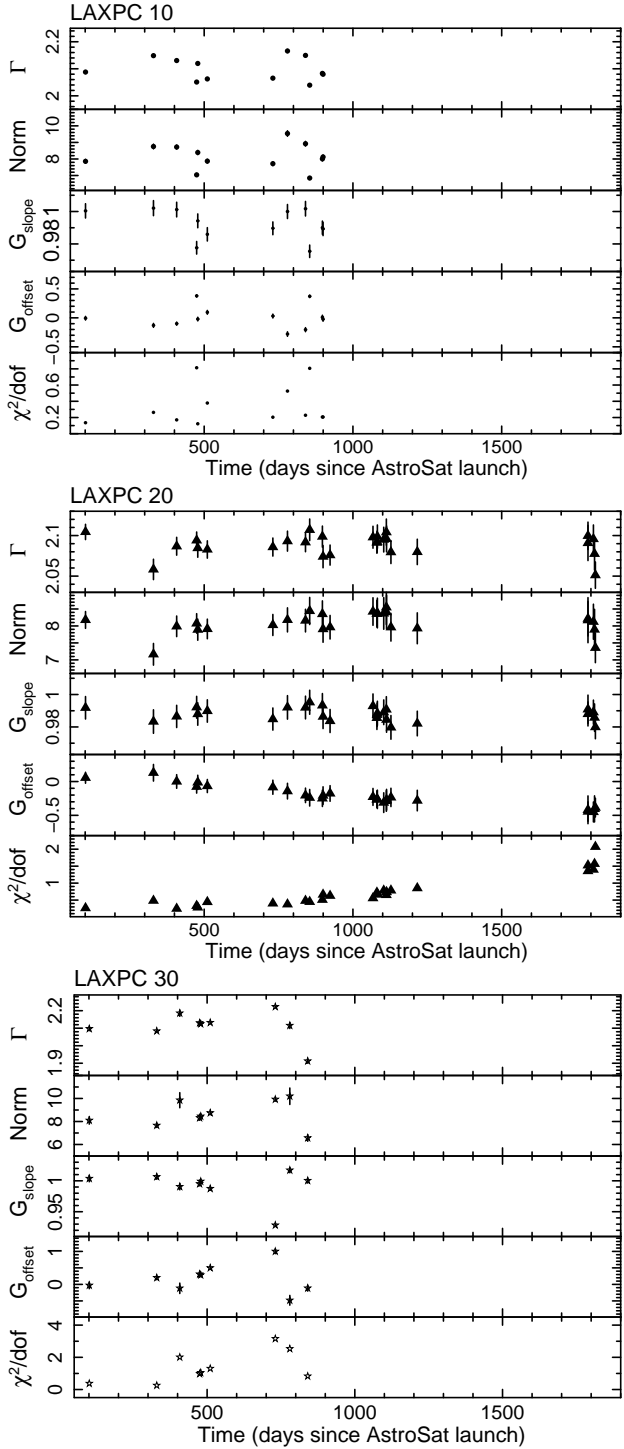

\centering
\includegraphics[height=1.00\columnwidth,angle=-90.0]{fig4a.eps}
\includegraphics[height=1.00\columnwidth,angle=-90.0]{fig4b.eps}
\includegraphics[height=1.00\columnwidth,angle=-90.0]{fig4c.eps}
\caption{The results of fit to Crab spectrum observed during the last 5 years is
shown as a function of days from launch of AstroSat for the three LAXPC detectors
as marked in the figure.  The different panels show all fitted parameters
as well as the reduced $\chi^2$ for the fit.}
\label{fig:crab}
\end{figure}
The energy resolution of LAXPC10 was stable until March 2018, while that for LAXPC30 was
improving with time, probably due to reduction in pressure.
The energy resolution of
LAXPC20 has been deteriorating with time and currently the resolution at 30 keV is about 20\%.
The position of peak channel in LAXPC20 has been steady and the last adjustment in HV was made 
on March 15, 2017. Since then the peak position has reduced by about 12 channels.
Since this detector is already operating at higher voltage (about 2600 V) as compared to the other two,
further increase in HV has not been attempted. Currently, this is the only detector
that is working nominally.

\subsection{Fit to Spectra of Crab Observations}

\begin{figure} [H]
\centering
\includegraphics[height=0.95\columnwidth,angle=-90.0]{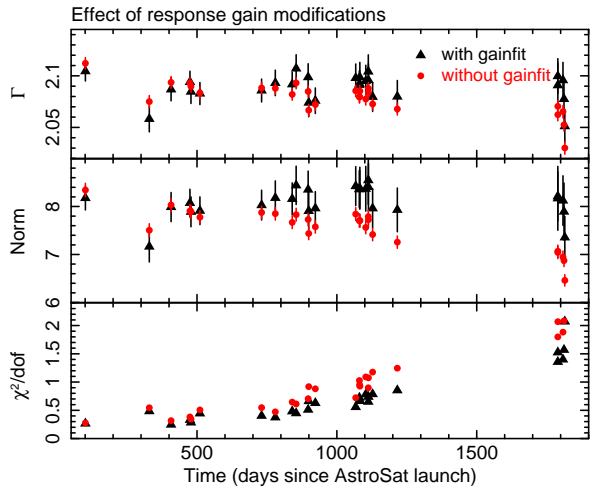}\quad
\caption{The results of fit to Crab spectrum observed by LAXPC20 during the last 5 years is
shown as a function of days from launch of AstroSat. The results with and without applying
the {\tt gain fit} are shown.}
\label{fig:crabgain}
\end{figure}

\begin{figure} [H]
\centering
\includegraphics[height=0.95\columnwidth,angle=-90.0]{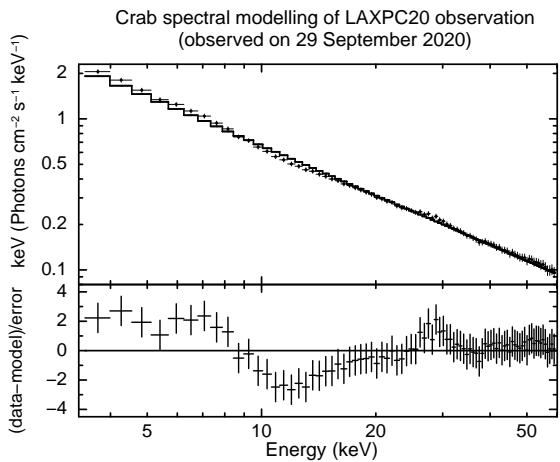}\quad
\caption{A fit to the Crab spectrum observed by LAXPC20 during September 2020 to powerlaw
	model with {\tt gain fit} is shown. The lower panel shows the residuals.
The reduced $\chi^2$ of the fit with 3\% systematics is 2.03.}
\label{fig:crabfit}
\end{figure}

AstroSat has observed the Crab X-ray source several times during the last five years
and the spectra obtained during these observations have been fitted to monitor the
stability of detector response after accounting for known drift in gain and pressure
using appropriate responses. The Crab spectra were fitted to a powerlaw form to obtain
the spectral index and normalisation for each observation (averaged over all orbits)
and the results are shown in
Figure~\ref{fig:crab}. All fits were performed with 3\% systematics in spectra and background,
and line-of-sight absorption column density of $0.34 \times 10^{22}$ cm$^{-2}$ was used
(Shaposhnikov et al.~2012).
Due to the anomalous behaviour and abnormal gain change, LAXPC10 and LAXPC30 data
were not fitted for 2018 onwards and hence are omitted from respective plots.
The {\tt gain fit} was also used in {\tt Xspec v 12.11.1} to allow for small deviations in the gain of responses.
The effect of using the {\tt gain fit} during the last five years for LAXPC20
is shown in Figure~\ref{fig:crabgain} where fitted spectral parameters with
and without using {\tt gain fit} are shown for comparison. 
It turns out that the
slope of best fit for LAXPC20 was always within 2\% of unity, which implies that this is largely
taken care of in gain shift estimated from the calibration source.
However, the offset in
{\tt gain fit} was found to change
systematically with time, reaching a value of $-0.5$ keV by now. This may be expected, as this
was not calculated from the calibration source. The inclusion of {\tt gain fit} improved the
fit significantly and is recommended for all spectral fits.
It can be seen that the fitted parameters have held steady
during the last 5 years, but the $\chi^2$ for the fits have increased with time.
Initially, 1\% systematics was enough to get an acceptable $\chi^2$ fit, but now it requires
up to 3\% systematics for LAXPC20. This could be because of degradation in energy
resolution. An example, of the fit for observation during September 2020 is shown
in Figure~\ref{fig:crabfit}
An estimate of systematic error in fitted parameter for Crab spectra can be obtained by
taking the standard deviation over all measurements. For the power-law index we get the
values, $2.099\pm0.041$, $2.088\pm0.013$ and $2.136\pm0.043$ for the three LAXPC detectors.
Similarly the normalisation is $8.15\pm0.74$, $8.10\pm0.31$ and $8.90\pm0.89$. Some of
the variation in normalisation could be due to differences in pointing offset over different
observations.

\begin{figure} [tbh]
\centering
\includegraphics[height=0.95\columnwidth,angle=-90.0]{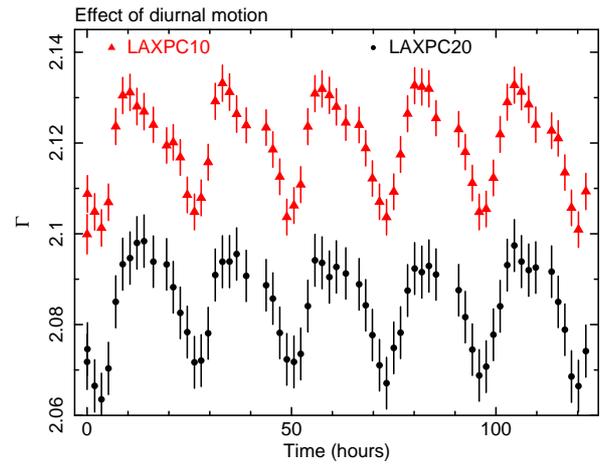}
\caption{An example of the diurnal variation observed in the
fitted Crab spectral parameters from LAXPC10 (shown by triangles) and LAXPC20 (shown by
circles) during one observation during January 2018 as a function of
time.}
\label{fig:crab24}
\end{figure}

\begin{figure} [tbh]
\centering
\includegraphics[height=0.95\columnwidth,angle=-90.0]{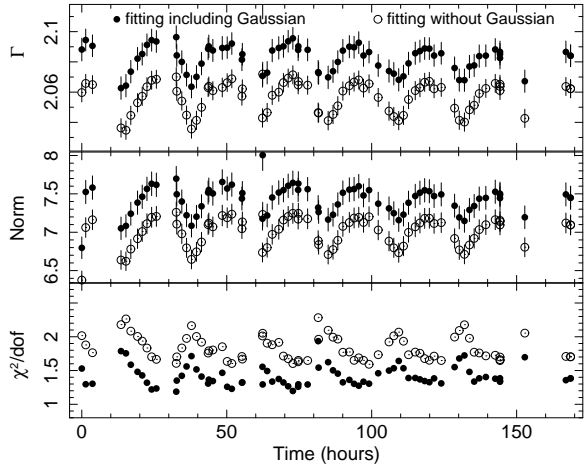}
\caption{ The figure shows the results of fitting the Crab spectra observed by LAXPC20
during a long observation during September 2020. Marginal improvement of
spectral parameters as well the fitting statistics can be noted when a powerlaw with a
Gaussian at Xe K X-ray energy $\sim30$ keV (shown by solid circles) is used compared to a simple
powerlaw (shown by hollow circles).}
\label{fig:crabgaus}
\end{figure}

It is clear that the fitted parameters for the Crab spectrum are stable over a long time scale.
Figure~\ref{fig:crab24} shows the variation over a short time scale of a few
days during January 2018 using data for individual orbits. It is clear that there is a diurnal variation in the fitted parameters, which is similar
to that seen in the background as shown in the next section. This variation is likely to
be due to variation in the background and discrepancy between the background estimated
from background model and the actual background.
Some of the variation could be due to a shift in GTI with orbit. The period of Earth
occultation would drift with time across the AstroSat orbital phase and that may account
for some of these variations. As a result some orbits will have more contribution from the
region near the SAA passage and these may have diurnal variation.
Since the LAXPC spectra often show an escape peak around 30 keV
due to Xe K X-rays, we also attempted a fit with an additional Gaussian peak around this energy
to account for this feature and the results are shown in the 
Figure~\ref{fig:crabgaus}. This resulted in some improvement in the fit and also reduced
the amplitude of diurnal variation, but significant variations were still seen.
The addition of a Gaussian component has been used in some analysis of LAXPC data to remove
the feature in the spectrum around 30 keV (Sreehari et al. 2019; Sridhar et al.~2019). 

\begin{figure} [tbh]
\centering
\includegraphics[height=0.95\columnwidth,angle=-90.0]{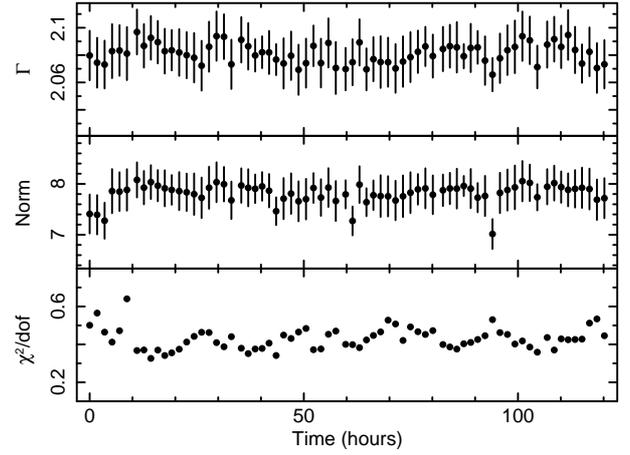}
\caption{The figure shows the results of fitting the Crab spectra observed by LAXPC20
during a long observation during January 2018 after removing the time-interval near the
SAA passage to reduce uncertainties in background estimates. The fit includes {\tt gain fit}
and a Gaussian around 30 keV.}
\label{fig:newgti}
\end{figure}

To identify the cause of the observed diurnal variation in the fitted parameters we repeated the
exercise for the January 2018 Crab observations, by removing the time intervals that
were within 600 s of entry or exit from SAA. This should reduce the background uncertainties.
With this modification the diurnal trend is not
clear as shown in Figure~\ref{fig:newgti}. These fits included the {\tt gain fit} as well
as a Gaussian around 30 keV and hence these results should be compared with those in
Figure~\ref{fig:crabgaus}.
However, with reduced exposure due to truncating the GTI, the errors in fitted
parameters are larger and the net range of fitted parameters is not substantially reduced.
It appears that some of the diurnal variation could be due to uncertainties in background
model discussed in the next section.
Although, Crab flux is much larger than the background at low energies, at high energies
it becomes comparable to background and hence can be affected by uncertainties in background.


\section{Detector Background}

To determine the detector background, the instrument is pointed to a direction where there
are no known X-ray sources (Antia et al.~2017). Since the background is found to show a
variation over a time of about 1 day, all background observations are for at least, 1 day
interval. The background counts are found to change during the orbit also with the counts
increasing near the SAA passage. These variations are
fitted to latitude and longitude of satellite as explained by Antia et al.~(2017).
To monitor the long term variation in the background, the background observations are
repeated about once every month. However, the variation in the background counts and
spectrum are too complex to be captured by the models and some of these complexities
are described in this section.

 \begin{figure*} [bht]
\centering
\includegraphics[width=0.95\columnwidth,angle=0.0]{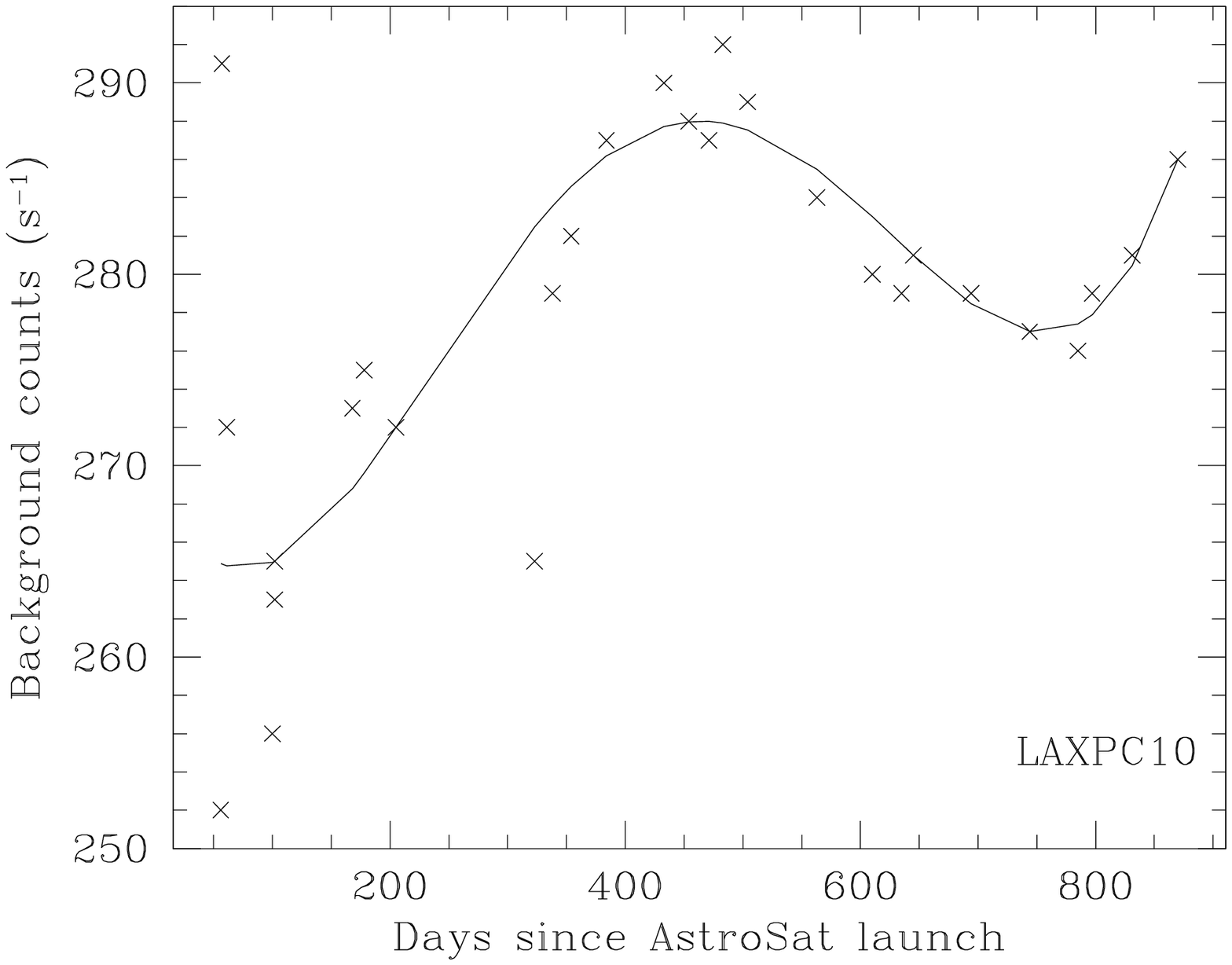}\quad
\includegraphics[width=0.95\columnwidth,angle=0.0]{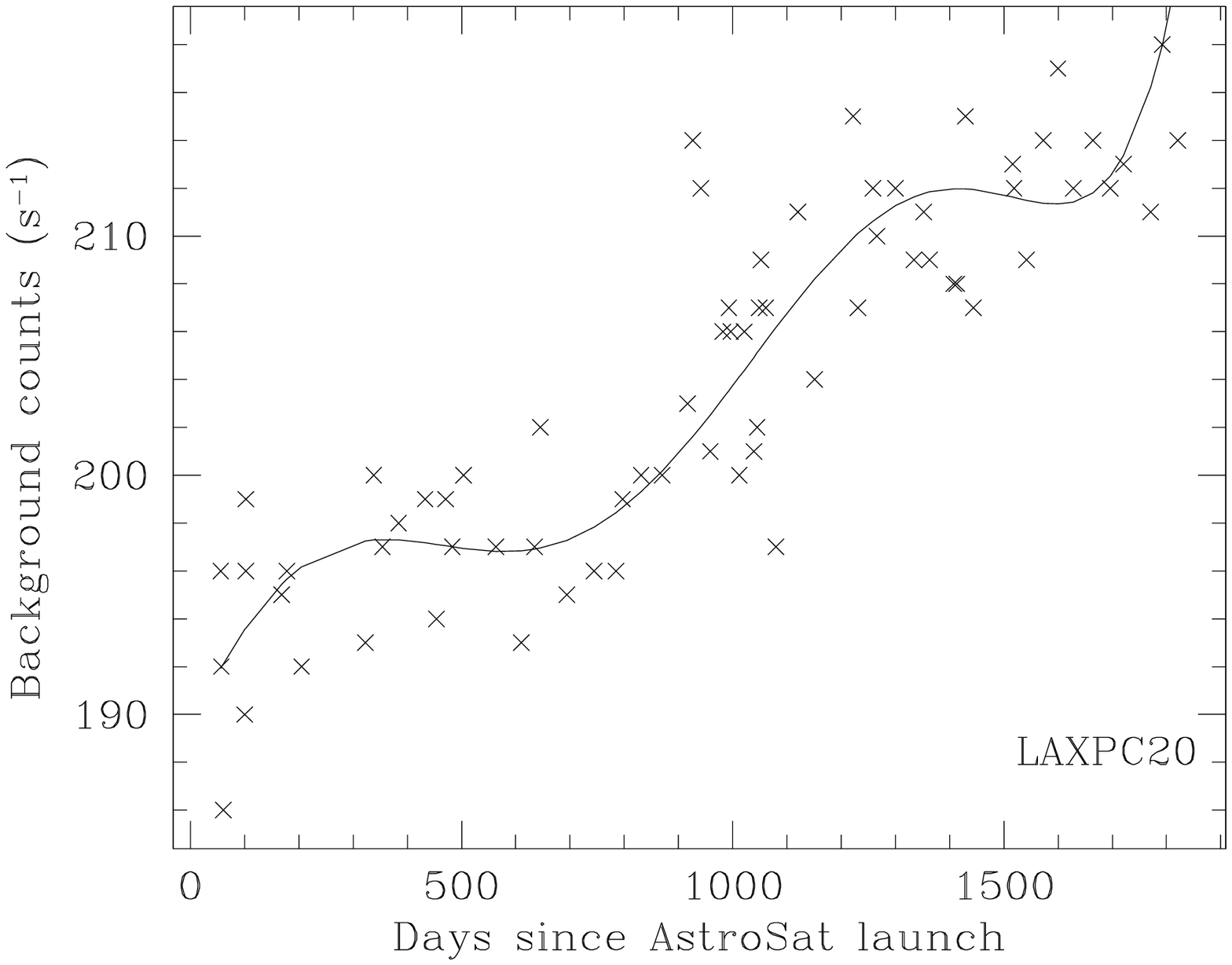}
\caption{The total count rate in background corrected for gain shift as a function of time
for LAXPC10 and LAXPC20.}
\label{fig:back}
\end{figure*}

The gain drift in detector would also result in change of background
and for spectral studies the background spectrum is corrected for this shift in the software. Even after correcting for
the gain shift, the background counts have been changing with time and the results are shown
in Figure~\ref{fig:back}. The LAXPC10 results are shown until March 2018 only, as after that
the gain has changed significantly. LAXPC30 results are not shown as the counts were decreasing
due to reduction in pressure and it is difficult to correct for large variations in the gain.
It can be seen that the variation is similar in both detectors during the overlapping time
and the counts have been generally increasing with time. The reason for this increase is not clear.
Some increase may be expected from induced radioactivity, but it is not clear why it becomes
nearly constant over some time intervals. There is also a significant scatter about the best
fit curve, which could be due to various factors discussed below.

\begin{figure} [th]
\centering
\includegraphics[width=0.95\columnwidth,angle=0.0]{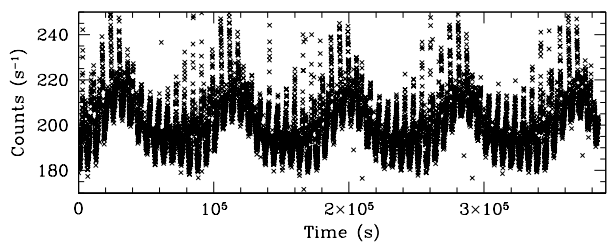}\quad
\caption{The light curve for a background observation during July 2018 in LAXPC20 with
a time-bin of 32 s.}
\label{fig:back2a}
\end{figure}

\begin{figure} [th]
\centering
\includegraphics[width=0.95\columnwidth,angle=0.0]{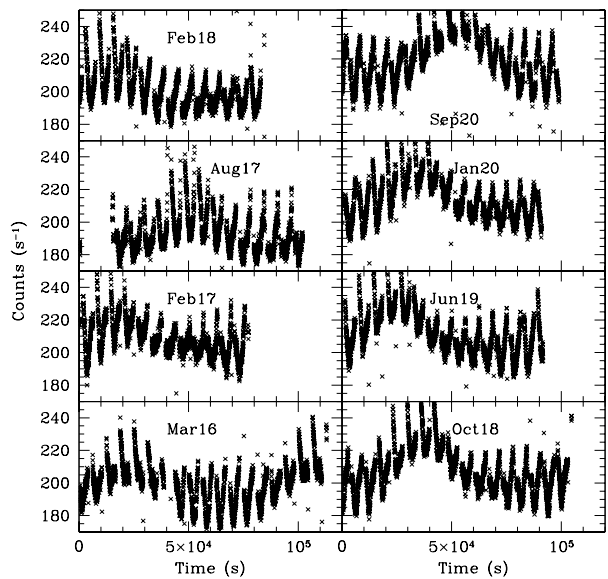}\quad
\caption{The light curves for  different background observations as marked with the month and year in the respective
panels, as observed by LAXPC20 with a time-bin of 32 s.}
\label{fig:back2}
\end{figure}

Figure~\ref{fig:back} shows the long term variation in the total count rate during
the background observations, but there are short term variations also during each orbit
as well as some diurnal variations during the course of a day. To show these variations
Figure~\ref{fig:back2a} shows the variation in the count rate for a long background
observation during July 2018. The diurnal variation can be seen clearly in this figure.
Figure~\ref{fig:back2} shows the light curves during a few background observations during
the last five years. It is clear that the diurnal variation is present in all observations
but there is some evidence that the amplitude of variation has increased with time.
Background model used to generate the light-curve for background has an option to remove
the diurnal variation with a period of 1 day, which can be applied if the observation
covers at least 1 day of observation. For short observations it tends to remove real
trend in the light-curve and hence is not applied. For longer observations also it can remove
some real variation if it happens to have similar periodicity.

 \begin{figure*} [th]
\centering
\includegraphics[width=0.95\columnwidth,angle=0.0]{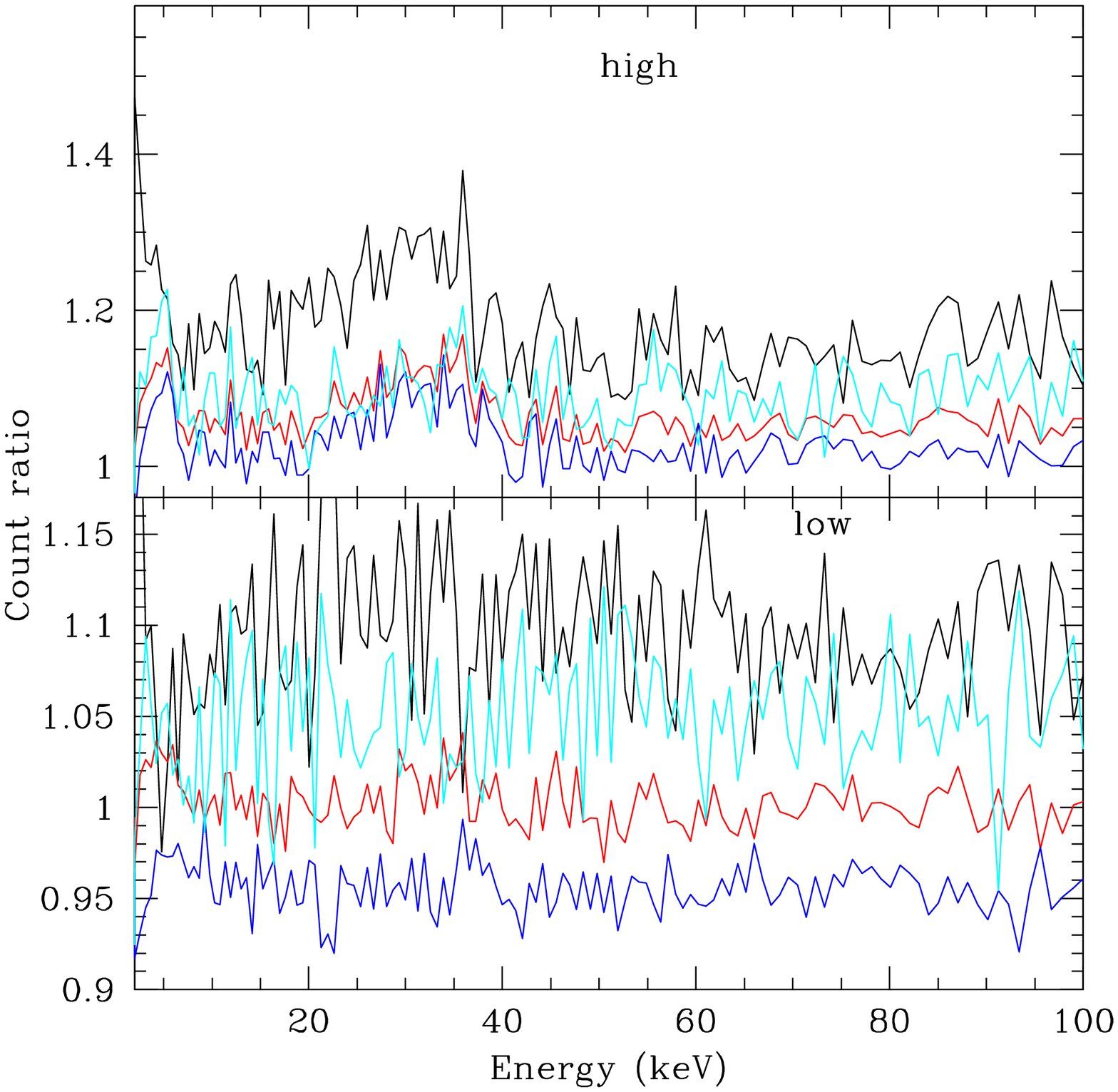}\quad
\includegraphics[width=0.95\columnwidth,angle=0.0]{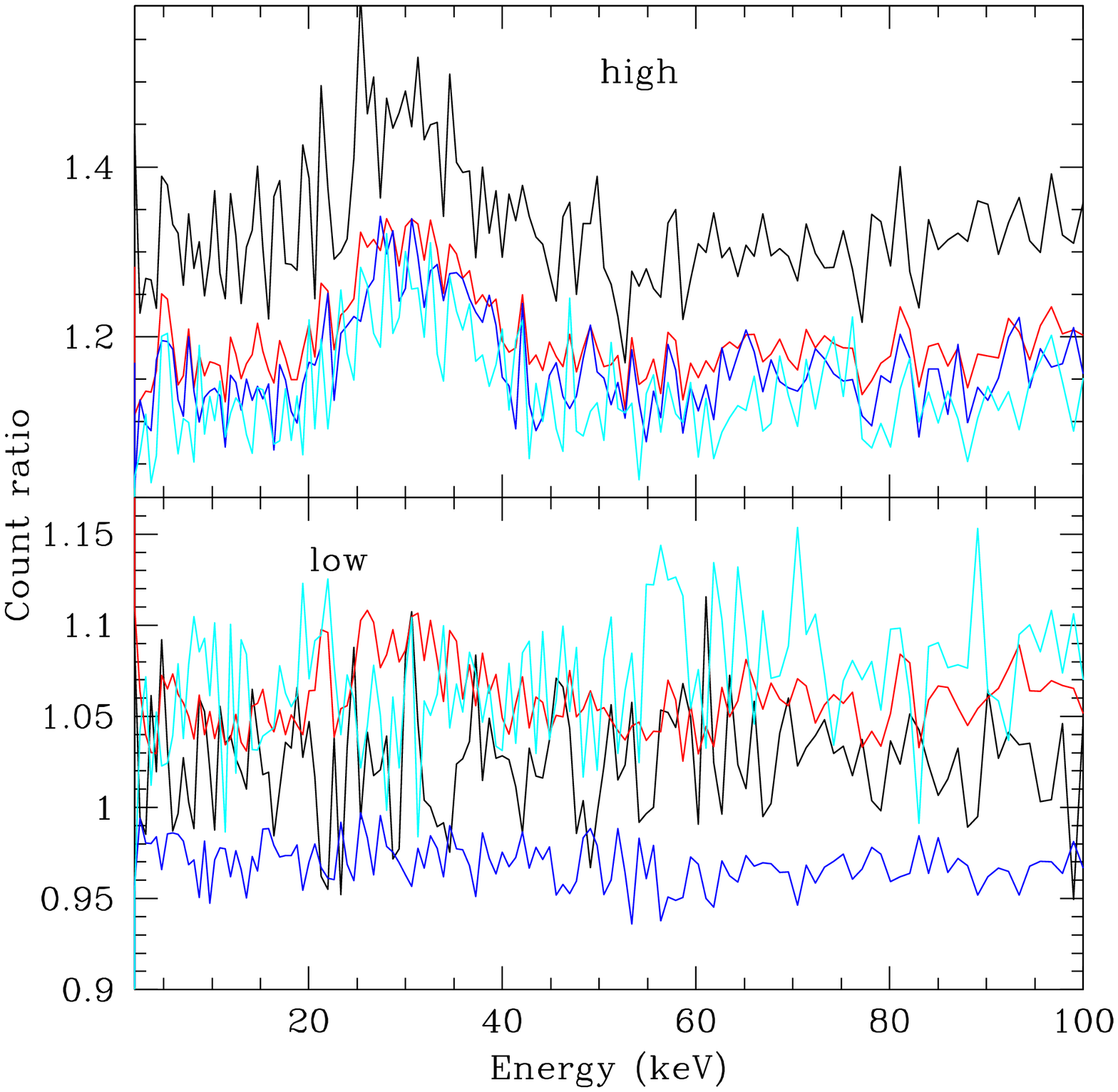}
\caption{The ratio of spectrum during different times of background observation with
respect to that during a `low' orbit for LAXPC20. The left panel shows the result for February 2017
background observation, while the right panel shows that for September 2020 observation.
The red line in top panels show the ratio averaged over `high' orbit, while that in the
bottom panel shows the ratio for entire observation, including all orbits. The black lines
show the ratio during the first 600 s of the orbit, the cyan line shows that during the
last 600 s, while the blue line shows that during the middle part of the orbit.}
\label{fig:back2p}
\end{figure*}

During an orbit the
counts are generally minimum in between the two SAA passages and tends to increase
as the satellite approaches the SAA, as well as when it exits the SAA. However, the
magnitude of variation near SAA passage is more complex. In general, for orbits where
the counts are near maximum in the diurnal trend, the counts have a sharp spike as
the satellite exits the SAA, while the spike when it approaches the SAA is of much
smaller magnitude. On the other hand, for orbits where the counts are near minimum of
the diurnal trend, the two branches, towards the entry to SAA and while exiting the
SAA are comparable in magnitude.
It is possible to avoid some of these artefacts by cutting off the interval surrounding
the SAA passage from GTI. However, the software does not implement this as it can reduce the
exposure time significantly, which may not be desirable in some studies, e.g., study of
bursts. For more sensitive studies it may be advisable to remove 600 s on either side of SAA
from the GTI.

All these figures only show the total background count rate, but the spectrum does not
simply scale by this rate and hence to look for the variation in the background spectrum, we selected the
background observations during February 2017 and September 2020 and calculated the spectrum during a few parts of
the observation. For reference, we use the spectrum obtained during an orbit when the
count rate was close to the minimum (referred to as `low') of diurnal variation, and choose another orbit when the count rate was close
to maximum (referred to as `high').
Here the orbit is defined as the period between two consecutive passages through SAA.
Figure~\ref{fig:back2p} shows the ratio of counts in the spectrum with
respect to the average spectrum during the `low' orbit.
The red curve in the top panel shows the ratio when spectrum is
averaged over the entire `high' orbit, while the red curve in the bottom panel shows
the ratio when the spectrum is averaged over the entire observation covering all orbits.
The black lines show the spectrum
during the first 600 s of the orbit just after the satellite comes out of SAA. The cyan line
shows the ratio when the spectrum is taken over the last 600 s before the satellite enters
the SAA, while the blue line shows the ratio for spectrum during the middle part of the
orbit, leaving out 900 s on both sides. The left panel shows the results for February 2017 observations, while the right panel
shows that during September 2020. It can be seen that during the `low' orbit
the difference is generally within 10\%.
However, during the `high' orbit all curves show a higher ratio and also
show a peak around 30 keV which is the Xe K X-ray energy.
Further, during the initial part of the orbit the flux is much higher at all energies with the
maximum difference exceeding 50\% for September 2020 observation. We have checked that even if we restrict to first
300 s of the orbit the ratio is only sightly higher.
Further, for `high' orbit the blue and cyan curves are close, indicating that there is
not much difference in the spectrum as the satellite enters the SAA. On the other hand,
for `low' orbit the counts increase as the satellite is about to enter the SAA region.
It turns out that the `high' orbits are the ones where
the passage through SAA occurs when the satellite is near the south end of its range.
Although, the result is not shown, the ratio of the spectrum during the two `low' orbits
is roughly consistent with the expected ratio from Figure~\ref{fig:back} except for a
peak around 30 keV. We have also looked at similar ratio using individual layers of the
detector and the behaviour is similar to that in Figure~\ref{fig:back2p}. However,
the count rate in the top layer is about twice that in other layers. Hence, the additional
counts are larger in the top layer as compared to other layers. Since the additional counts
are larger after exit from SAA, some of these could be due to induced radioactivity, while
there could be additional contribution from charged particles coming through the collimator.
Thus it is clear that there is a significant change in the background spectrum during later
times and most of the increase in background appears to be during `high' orbits and
for energy around 30 keV. 

\begin{figure*} [thp]
\centering
\includegraphics[width=0.95\columnwidth,angle=0.0]{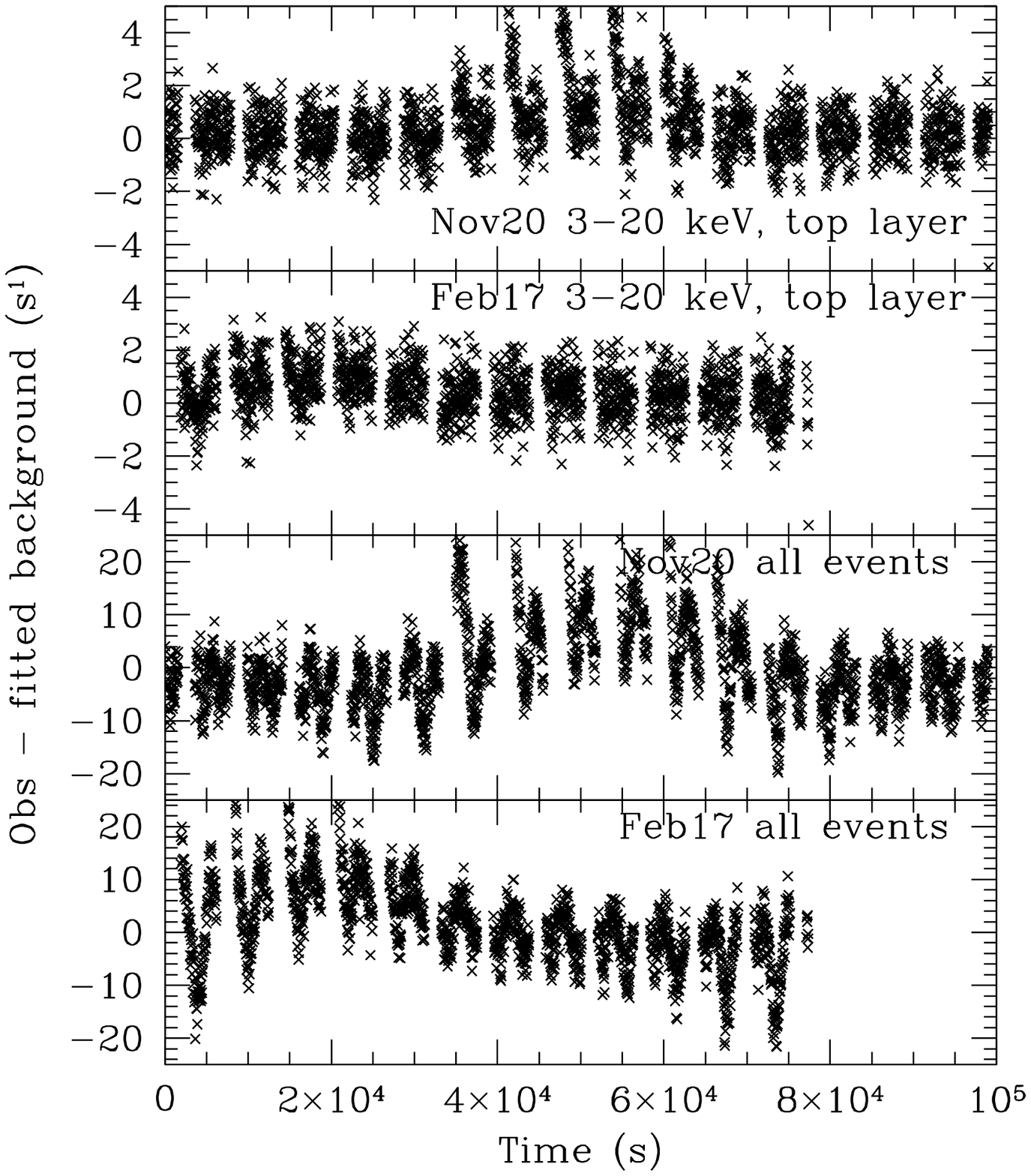}\quad
\includegraphics[width=0.95\columnwidth,angle=0.0]{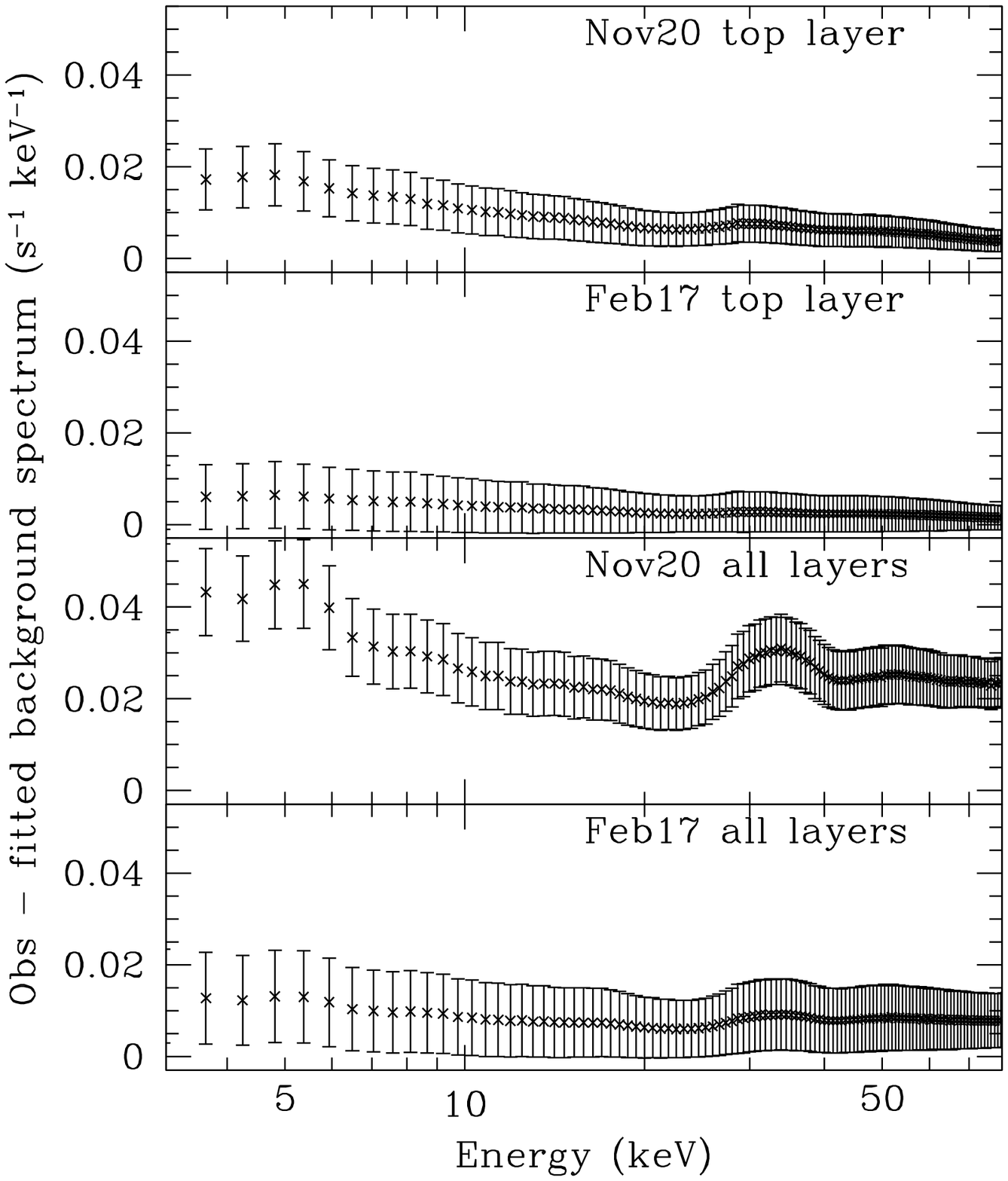}
\caption{The residuals in the fit to background of LAXPC20 for the two background observations
obtained using the background model described by Antia et al.~(2017). The left panel
shows the residuals in the light curve with a time-bin of 32 s, while the right panel
shows the residuals in the energy spectrum.}
\label{fig:backfit}

\centering
\includegraphics[height=0.95\columnwidth,angle=-90.0]{fig15a.eps}\quad
\includegraphics[height=0.95\columnwidth,angle=-90.0]{fig15b.eps}
\includegraphics[height=0.95\columnwidth,angle=-90.0]{fig15c.eps}\quad
\includegraphics[height=0.95\columnwidth,angle=-90.0]{fig15d.eps}
\caption{The residuals in the fit to background for the two background observations
obtained using the background model for faint sources described by Misra et al.~(2020)
which uses only the top layer of the detector. The top panels
shows the residuals in the light curve with a time-bin of 100 s, while the bottom panels
shows the residuals in the energy spectrum. The left panels show the results for February 2017
background observation while the right panels show the same for September 2020
observations.}
\label{fig:backalt}
\end{figure*}

The background model used in the software does account for the increase in the count rate
during the `high' orbit to a large extent, but the spectrum is scaled to the average
counts and hence is likely to introduce a bump around 30 keV. Typical rms deviations in
the fit to count rate are 10 s$^{-1}$ when total counts in all anodes are considered.
For top layer in 3--20 keV energy range this drops to about 1 s$^{-1}$, which is
comparable to statistical error for a time-bin of 32 s, used in these fits.
This may be expected as the deviations are more prominent in energies above 20 keV.
Figure \ref{fig:backfit} shows the residuals in the background fit for the two
observations described above using the background model described by Antia et al.~(2017).
It can be seen that for the restricted energy range using
only top layer, the residuals are roughly consistent with the statistical errors, except for
the high orbits during November 2020 observations. However, when all events are considered
the background model has significant residuals. Thus it is clear that for faint sources it
would be advisable to consider only top layer of the detector with restricted energy
range. Figure \ref{fig:backalt} shows the residuals in the background fit for the same two
observations using the background model for faint sources described by Misra et al.~(2020)
which uses only the top layer of the detector.

\begin{figure} [th]
\centering
\includegraphics[width=0.80\columnwidth,angle=0.0]{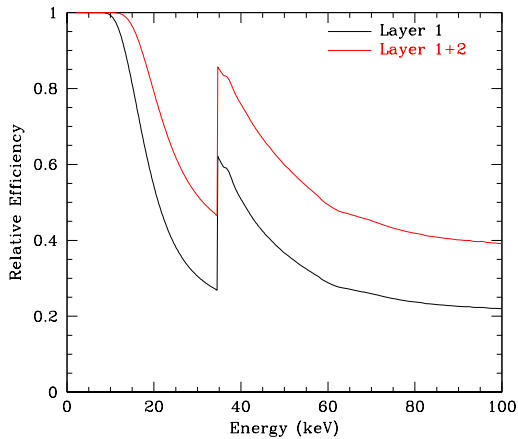}
\caption{The relative efficiency of only top layer and the top two layers of
LAXPC detector as compared to when all layers are used.}
\label{fig:lay}
\end{figure}

Since the background model performs better when only top layer of the detector at low
energies is used, it is advisable to use this option for faint sources. This is
justified as at low energies a large fraction of counts are registered in the top layers.
Figure~\ref{fig:lay} shows the energy dependence of relative fraction of events in top layer
(Anodes 1 and 2) and
the top two layers (Anodes 1 to 4) as compared to all layers (Anodes 1 to 7).
It can be seen that up to about 10 keV, almost all events are registered in the top layer.
Even at 20 keV about 50\% of events are registered in the top layer and about 75\% in the
first two layers.
Thus for studies at low energies it is advisable to use only the top layer of the detector,
as it reduces the background.
An alternative is to
drop the data during the `high' orbits to avoid this contamination. This would reduce the
duty cycle of the observations and if the observation is for a short duration, the entire
part may be during the `high' orbits. The fluctuations in the background determine the
sensitivity of LAXPC for faint sources, as discussed in the next section.

 \begin{figure} [th]
\centering
\includegraphics[width=0.48\columnwidth,angle=0.0]{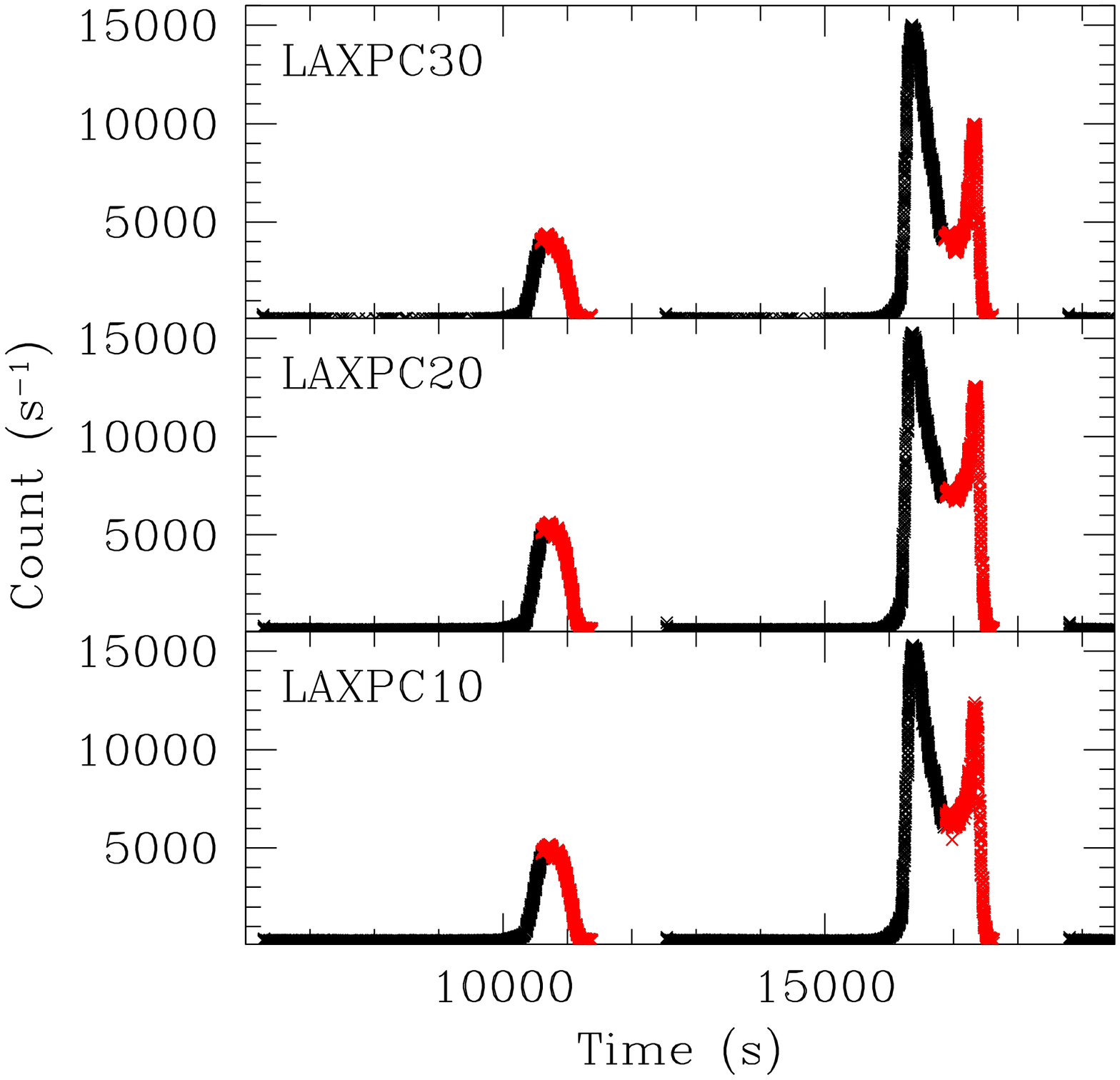}\;
\includegraphics[width=0.48\columnwidth,angle=0.0]{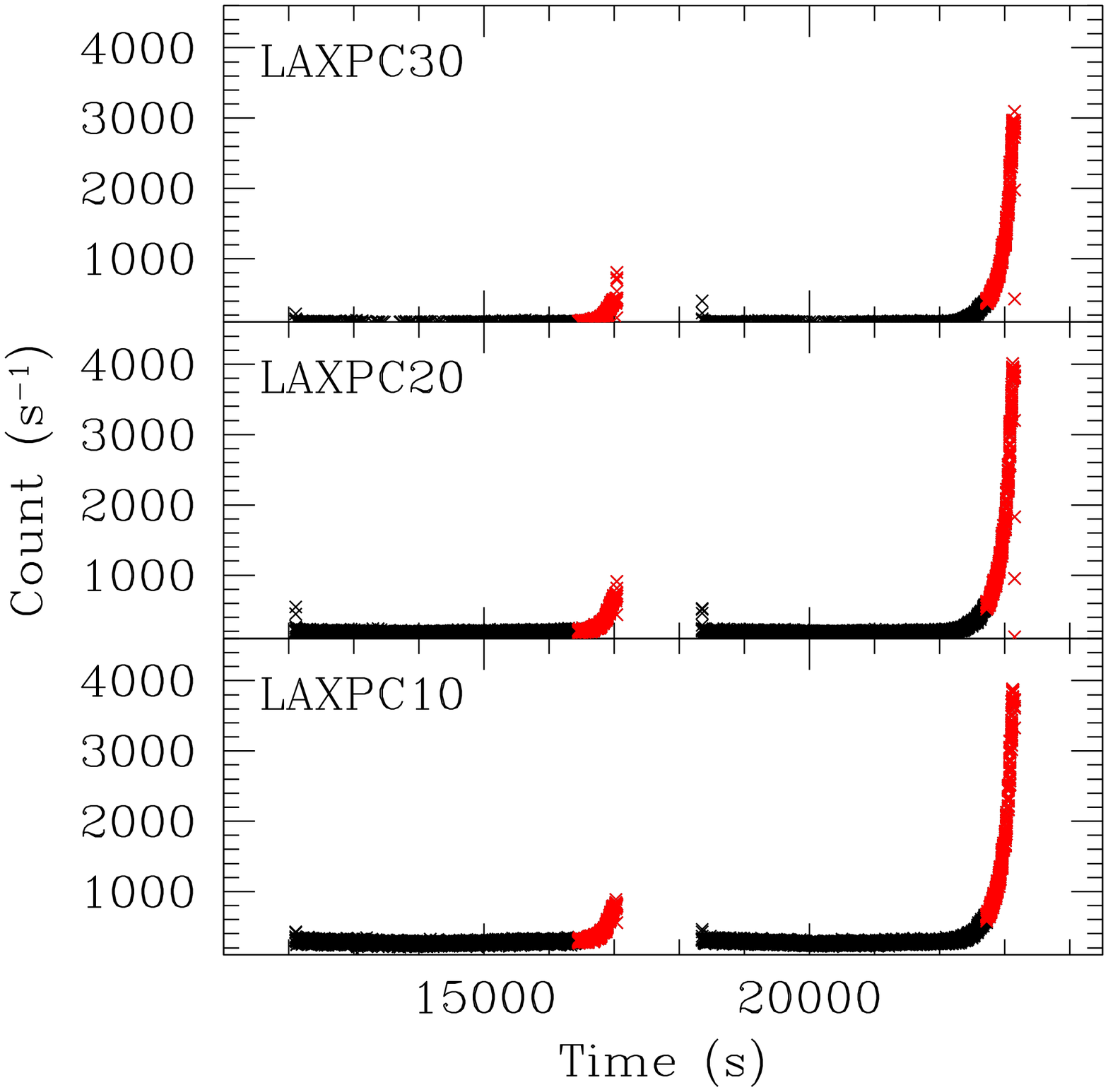}
\caption{The light curve during geomagnetic storms on 
on September 8, 2017 (left panel) and May 28, 2017 (right panel).
The red points show the part of the light curve which would be outside the GTI.}
\label{fig:kp}
\end{figure}

Since the detector background increases close to SAA passage, the contribution is most
likely from the flux of charged particles. Although, the detector has veto anode and shield on three
sides to protect from charged particles entering from these sides. On the two smaller
faces there is only a shield which would offer some protection. However, on the top side there
is no veto layer or shield, unlike that in RXTE/PCA which had a propane layer (Jahoda et al.~2006). As a result, there
is little protection other than a thin Mylar window, against charged particle coming from the top along the collimator.
The opening angle of collimator is about $1^\circ$ which gives a solid angle of about 0.0003 st.
Thus even for a flux of 1 particle cm$^{-2}$ s$^{-1}$ st$^{-1}$, the detector with an effective
area of 2000 cm$^2$ would record 0.6 event per s. This level of flux is entirely possible even
outside SAA, while close to SAA the flux could be larger.

During a geomagnetic storm the charged particle flux goes up and many
more counts are recorded. Figure~\ref{fig:kp} shows the count rate in detectors during
two geomagnetic storms on September 8, 2017 ($Kp=8^+$) and May 28, 2017 ($Kp=7$).
The $Kp$ index in the parentheses gives a measure of strength of geomagnetic storm. The September 8, 2017
event was the strongest so far during the AstroSat operation.
The red points marks the time which would be outside the GTI due to normal SAA definition
and would not be considered for analysis. Thus it can be seen that geomagnetic storms
can yield significant counts near SAA passage. The most affected regions in this
case are those where SAA passage occurs during the north end of the range. For weaker
storms, the effect will not be easily seen in light-curve as the net increase would
be smaller but such storms can be frequent during high activity part of the solar cycle.
Even for smaller geomagnetic storms, of the order of 10 c s$^{-1}$ can be added before the SAA
passages. If such an event occurs during observation of a variable source, it would be almost
impossible to separate it out from variation in source counts. Any burst seen close to
SAA passage can be suspect and would require further investigation.
During the last two years the solar activity has been low and such storms are
rare, but in the coming year the solar activity is likely to pick up and more such events
would be seen.

\section{Detector Sensitivity and Limitations}

There is no difficulty in studying bright sources. The flux limit for faint sources
is determined by the fluctuations in the detector background and ability of background
model to match these variations. Even for long observations
where the spectrum is averaged over a long time, the intrinsic variations in background
cannot be modelled satisfactorily and it limits the sensitivity of the detector for faint sources.
From the discussion in the previous section we can see that at least, a few counts per
second from the source would be required to get any meaningful results. The actual limit
would obviously depend on the level of details that need to be studied and the fluctuation
in background during actual observation. The Crab observation yields a count rate of
about 3000 s$^{-1}$ in each detector, which gives a sensitivity limit of about 1 mCrab for faint sources
that can be studied. For reference, low energy flux of about $10^{-11}$ erg cm$^{-2}$ s$^{-1}$ gives
a count rate of 1 s$^{-1}$. The same difficulty arises even for relatively bright sources
at high energies where the count rate can be much less than that in the background.
The upper limit on energy to which the
spectrum can be studied depends on the source and the extent of details that are required.
In the following subsections we illustrate some limitations and
capabilities of LAXPC for studying different properties of X-ray sources.
Instead of giving limit on flux etc., we illustrate the capabilities and limitations by giving some examples.
All the results presented in this section are obtained using the software described in
Section 5.2 with default parameters, except for choice of energy range and anodes as specified.

 \begin{figure} [th]
\centering
\includegraphics[height=0.95\columnwidth,angle=-90.0]{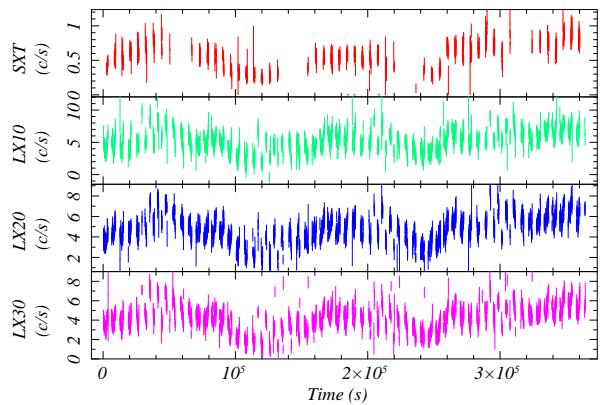}
\caption{The light curve of NGC 4593 during the AstroSat
observation starting on July 14, 2017 in SXT and the three LAXPC detectors with a time-bin
of 100 s.}
\label{fig:ngc}
\end{figure}

 \begin{figure} [th]
\centering
\includegraphics[height=0.95\columnwidth,angle=-90.0]{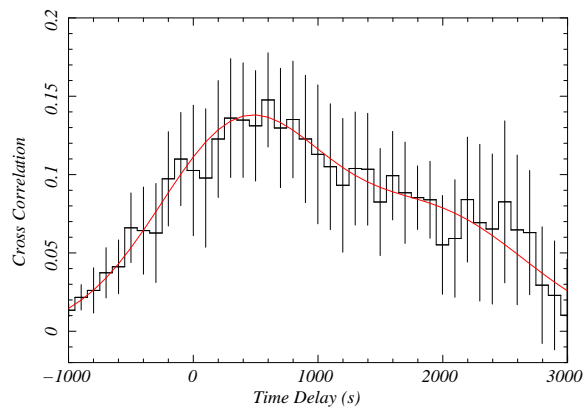}
\caption{The Cross-correlation between SXT and LAXPC20 light-curve of NGC 4593 is shown as a
function of time delay. The red line shows the fit with 2 Gaussians.}
\label{fig:ngcor}
\end{figure}

 \begin{figure} [th]
\centering
\includegraphics[height=1.00\columnwidth,angle=-90.0]{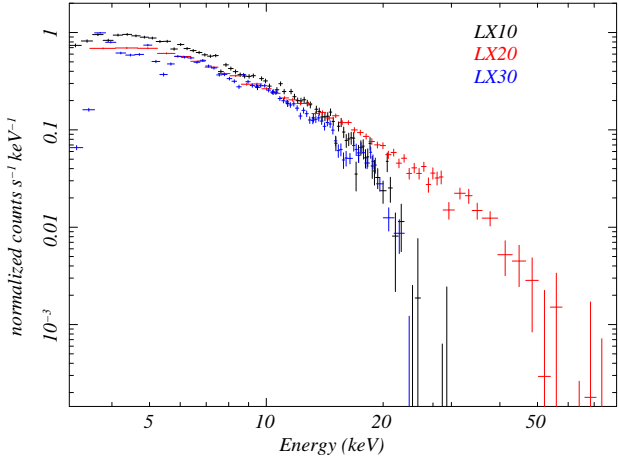}
\caption{The spectrum of NGC 4593 during the AstroSat
observation starting on July 14, 2017 in the three LAXPC detectors.}
\label{fig:ngcspec}
\end{figure}

 \begin{figure} [th]
\centering
\includegraphics[height=1.00\columnwidth,angle=-90.0]{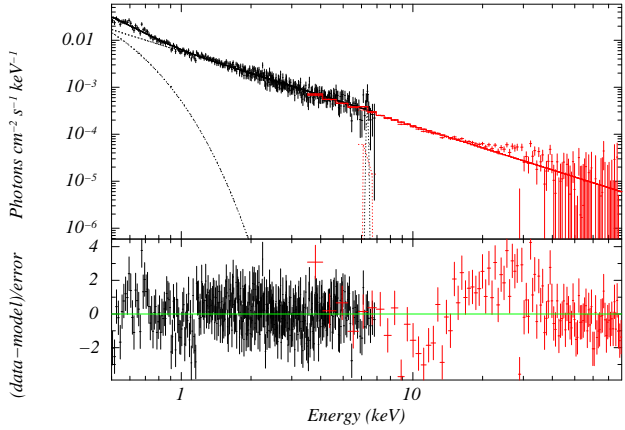}
\caption{The fit to the combined spectrum from SXT and LAXPC20 for NGC 4593 is shown
with the bottom panel showing the residuals.}
\label{fig:ngcfit}
\end{figure}

\subsection{Light Curve and Spectrum}

For faint sources typically most flux is observed at low energies and it is better
to use only the top-layer of the detector and further restrict the energy range
to 3--20 keV. This reduces the background counts by a factor of 10, thus improving
the signal to noise ratio.
To illustrate the limitation we have selected an AstroSat observation of an Active Galactic
Nuclei (AGN), NGC 4593, on July 14, 2016\footnote{ID: 20160714\_G05\_219T01\_9000000540}.
Figure~\ref{fig:ngc} shows the LAXPC light-curve using only top layer and
energy range of 3--20 keV. This source shows a count rate of about 5 s$^{-1}$ in all
three detectors. For comparison the light curve at low energies from the SXT instrument
is also shown. All light curves are with a time-bin of 100 s to reduce the statistical error.
It can be seen that all LAXPC detectors and the SXT instrument show similar variation.
The SXT being an imaging instrument has low background and further it operates at lower
energy of 0.3--10 keV, where the flux is larger. Hence, it is expected to give reliable
light-curve for this source. The cross correlation function {\tt crosscor} of {\tt HEASoft v 6.28} tool was
used to estimate the time lag between the LAXPC20 (3--20 keV) and SXT (0.5--3.0) keV which
is shown in Figure~\ref{fig:ngcor}.
A standard plot of
cross-correlation versus time delay was generated using these energy bands with time resolution of 99.86 s
with 512 intervals. Two Gaussian models were fitted to the cross-correlation to evaluate the time lag. The
resulting fit shows a clear time delay of  about 400~s indicating that hard photons lag behind the soft
photons (Brenneman et al.~2007).

Figure~\ref{fig:ngcspec} shows the spectrum observed in each
of the LAXPC detectors. There is a reasonable agreement between the three detectors at low energies.
The LAXPC20 spectrum continues at high energy also, probably because the background is
more reliably estimated. To test the spectrum, the combined spectrum from SXT and LAXPC20
covering 0.5--80 keV was fitted using {\tt Xspec} to the model phabs*(diskbb + gaussian + powerlaw)
and the resulting fit is shown in Figure~\ref{fig:ngcfit}.
The model fit yielded a reduced chi-squared of 1.45
(639/442) with 1.5\% systematics. Also, significant residuals were seen in the 10--30 keV range, which might be due to a
broad reflection component. The disk temperature ($T_{in}$) and photon index ($\Gamma$) obtained were $0.13\pm 0.01$ keV
and $1.56\pm 0.01$, respectively,  which are consistent with the results of Ursini et al.~(2016).

\subsection{Pulsation}

\begin{figure} [th]
\centering
\includegraphics[width=0.95\columnwidth,angle=0.0]{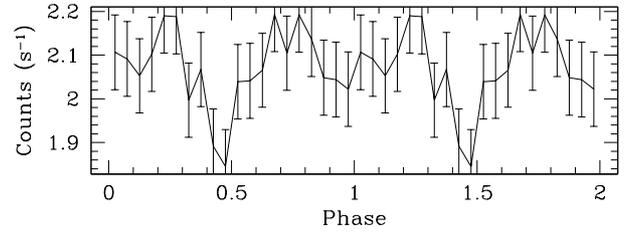}
\caption{The pulse profile of SMC X-2 as obtained by LAXPC20 in the energy range 3--20 keV.}
\label{fig:smcx2}
\end{figure}

Several X-ray pulsars have been studied by LAXPC and in general there is no difficulty
in estimating the frequency and spin-up rate if the change in frequency is significant.
To illustrate the performance, we consider the AstroSat observation of SMC X-2 on
May 7, 2020 (ID 20200507\_T03\_205T01\_9000003652), when the average count rate from the source was 2.1 s$^{-1}$.
Nevertheless, it was possible to estimate the spin period of $2.377441\pm0.000016$ s at the beginning
of observation (MJD 58976.575405) and the spin-up rate of $(3.9\pm 1.1)\times10^{-11}$ Hz s$^{-1}$
using LAXPC20 in single event mode and energy range of 3--20 keV.
The resulting pulse profile shown in Figure~\ref{fig:smcx2} can be compared
with other observations (Li et al.~2016; Jaiswal \& Naik 2017). Thus, it is clear that even
with a low count rate it is possible to study pulsations.

Apart from coherent pulsation, LAXPC has been extensively used to study Quasi Periodic Oscillations (QPO)
in a wide variety of sources, with frequency ranging from 1 mHz in 4U 0115+63 (Roy et al.~2019)
to 815 Hz in 4U 1907+09 (Verdhan Chauhan et al.~2017). The only problem with detecting QPO
arises if their frequency is close to the orbital frequency of AstroSat ($\sim 0.15$ mHz)
or its harmonics. Up to 10 harmonics of orbital frequency can be easily seen in the
power density spectrum and need to be accounted for while identifying QPO frequencies.
Similarly, for X-ray pulsars there can be interference from pulse frequencies or its harmonics.
But this can be easily removed by modelling the pulse including harmonics and removing their
contribution, e.g., for GRO J2058+42 (Mukerjee et al.~2020a).

\subsection{Cyclotron Resonant Scattering Features}

Detection  and studies of Cyclotron Resonant Scattering Features (CRSF) has always been  of
great interest for
direct measurements of magnetic field of the neutron stars and  to understand  structure of
the line forming regions
in the accretion column in X-ray binary pulsars. The LAXPC instrument was designed  to study
CRSF in X-ray pulsars
in binaries, as one of its important defined objectives with superior detection efficiency
along with its high time
resolution capabilities covering  3--80 keV wide energy band. Detection of CRSF requires
accurate  spectral response
and a well defined continuum model which enables  detection of absorption features  in the
source spectrum due to
CRSF. For a genuine detection  of  absorption features  in the spectrum and to minimise the
possibility  of false
detection  due to any unknown  discrepancy  of spectral response over entire energy band, one
can cross verify and
re-confirm  presence of  these absorption features indirectly by  comparing  the ratio of
source  spectrum to that
of the well known Crab spectrum having well calibrated power-law spectrum (e.g., Mukerjee et
al.~2020a). One can
also quantify energy non-linearity, if present, in the spectral response by analysing the Crab
ratio plot from known
CRSF sources. The LAXPC spectral response were generated carefully by appropriately modelling
and accounting for
30 keV florescence photons  produced due to Xe K shell interaction of incident X-rays.
However, some feature around 30 keV is still seen in most spectra which can interfere with
detection of CRSF.
Nevertheless, LAXPC, has detected CRSF in many
X-ray pulsars such as in GRO J2058+42 (Mukerjee et al.~2020a), Her X-1 (Bala et al.~2020) and
Cep X-4 (Mukerjee et al.~2020b) with energies in 30--40 keV region.
As an example of
detection capability of LAXPC at the highest energy, we attempt to study CRSF in GRO J1008--57,
where such feature
has been detected at 90 keV (Ge et al.~2020). We used AstroSat observation of this source on
November 7, 2017\footnote{ID: 20171107\_A04\_024T04\_9000001670} with
exposure time of 43 ks to investigate CRSF at such a high
energy close to the ULD of LAXPC20.

 \begin{figure} [th]
\centering
\includegraphics[height=0.90\columnwidth,angle=-90.0]{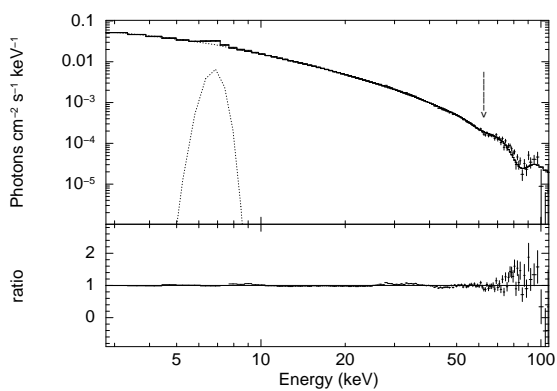}
\caption{The fit to LAXPC20 spectrum of GRO J1008--57 including CRSF at 61 and 84.4 keV with
the ratio of the data and fitted model shown in the lower panel. The position of 61 keV
feature is marked by an arrow.}
\label{fig:crsf}
\end{figure}

Figure~\ref{fig:crsf} shows a fit to the  GRO J1008-57 spectrum derived from AstroSat
observation.  The spectral data
was reasonably defined by the combined model defined as
phabs(gaussian+powerlaw)highecut*2gabs. The ratio of the data
and the fitted model is also shown below for clarity. The power-law with high energy  cut-off
model defines the
continuum  reasonably well with parameter  values derived which is typical for  an X-ray
binary pulsar.
The hydrogen column density value was frozen at $n_H=1.22 \times 10^{22}$ cm$^{-2}$ as calculated using
{\tt HEASARC} tool for
the source.  The well known Fe-line emission was detected  and its energy and width were fixed
during the fit at 6.5 keV and 0.3 keV, respectively.
The photon-index was found to be $1.13^{+0.40}_{-0.07}$, E-cutoff at $7.96^{+0.75}_{-0.83}$ keV and
E-fold at
$26.65^{+0.78}_{-0.79}$ keV. The centroid energy of CRSF was detected  at
$84.4^{+7.3}_{-4.0}$ keV
with the line-depth of $10.1^{+5.2}_{-4.4}$, while the width was frozen at 5.5 keV to constrain
its upper limit within ULD.
The reduced $\chi^2$ of the model fit was  1.6 for 128 degrees of freedom. A systematic
error of 2.5\% was added
to account for uncertainties in the response.  The CRSF is thus  detected with a significance
of about $3\sigma$.
Even though this energy is near the upper limit of the nominal range of LAXPC, it turns out that
the ULD in LAXPC20 is
around 100 keV and hence an absorption feature  could be  seen in the spectrum around 85 keV.
The presence of this
feature was also cross verified in the spectral ratio with Crab spectrum. The detected  CRSF
energy 
is within the error limit of that reported by Ge et al.~(2020).
Interestingly, in the AstroSat spectrum, an additional absorption feature was also detected
around $61.2^{+1.7}_{-1.5}$ keV which has not been reported earlier.

\subsection{Source Contamination}

\begin{figure} [th]
\centering
\includegraphics[width=0.90\columnwidth,angle=0.0]{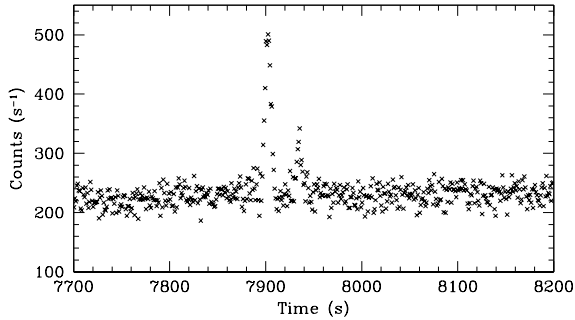}
\caption{The light curve during a slew operation on May 7, 2019.
The two peaks in the light curve are due to GX 5--1 (offset $51'$) and
GX 9+1 (offset $62'$).}
\label{fig:slew}
\end{figure}

A major problem with LAXPC is its large field of view, with FWHM of about $1^\circ$ (Antia
et al.~2017), which allows multiple sources in the field of view. If the contaminating
source has an angular offset of $30'$, then its flux would be reduced by about half, which
can give significant contamination depending on relative flux from the two sources.
Even at an offset of $1^\circ$, 5\% of the flux may be registered in LAXPC instrument,
 a part of which may be attributed to the leakage of higher energy ($>50$ keV) photons through the
collimator.
A test of this is provided by slew observations, where many known sources are seen through
a bump in the light curve. There have been cases when the source was clearly visible even when
the offset was $65'$. For example, during the slew observation\footnote{ID: 202190507\_SLEW\_01234\_9000002893} on May 7, 2019, the light-curve in LAXPC20 (Figure~\ref{fig:slew})
shows two clear peaks. The first peak with additional counts of 300 s$^{-1}$ is due to
GX 5--1 which had an offset of $51'$. This source has been observed several times by
AstroSat and typical count rate is about 3000 s$^{-1}$, thus it appears that about 10\% of
flux is registered at this offset. The second peak with height of about 100 s$^{-1}$ is
due to GX 9+1 which had an offset of $62'$. This source has also been observed twice by
AstroSat with a count rate of about 1800 s$^{-1}$. Thus even at this offset about 5\%
of flux is registered.
Hence it is recommended
that ideally there should be no significant source within $75'$ of the target source.

Some well known examples of sources affected by contamination are GRS 1758--258, 4U 1630--472 and
IGR J17091-3624. For GRS 1758--258 there is another source, GX 5--1, $40'$ away, which is more
than 10 times brighter. As a result, the contaminating source overwhelms the target.
Even if the observation is made with an offset of $30'$ on the opposite side, GX 5--1
can still contribute 5\% of its on-axis flux which would be comparable to 50\% of
the GRS 1758--258 flux. Thus it is difficult to study this source using LAXPC. Further, with such
offset the target would be outside the field of view of the SXT
and hence it would not be possible to do broadband spectral studies. Similarly,
4U 1630--472 has contamination from AX J1631.9--4752 which is $35'$ away. In this case,
both sources have comparable flux and it may be possible to subtract the contribution
from contaminating source (Baby et al.~2020). The 1310 s pulsation from the contaminating
source were used to estimate its flux. There are a few other sources also in the
field of view, but they are probably transients and may not be in outburst at the time of
observation. But this needs to be checked. For IGR J17091--3624, there is a contamination
from GX 349+2, $40'$ away, which could be a few times brighter than the target.
It may be possible to remove the contamination, if simultaneous observation of the
contaminating source are available from other instruments, or an estimate of flux
is available from monitoring instruments like, MAXI or Swift/BAT (Katoch et al.~2020).

 \begin{figure} [th]
\centering
\includegraphics[width=0.80\columnwidth,angle=0.0]{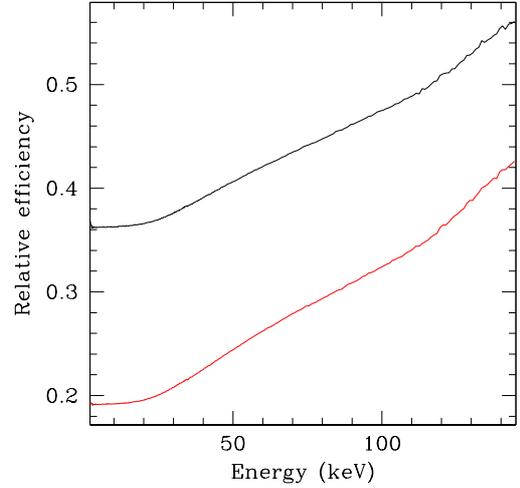}
\caption{The relative efficiency of LAXPC20 detector with offset of $30'$ (black line)
and $40'$ (red line) as compared to on-axis response.}
\label{fig:off}
\end{figure}

To remove the contamination, the contaminating source needs to be modelled and a
response with offset needs to be used to calculate its contribution to the observed
spectrum. The responses with offset have been calculated for LAXPC20. Figure~\ref{fig:off}
shows the relative efficiency of LAXPC20 for offset of $30'$ and $40'$ with respect to
on axis response. It can be seen that although at low energy the efficiency can be
somewhat low, it increases with energy. These responses have been averaged over
a circle with a given offset. There would be some dependence on the angle with respect to
detector side. The efficiency would be higher when the source is at the same offset
along the diagonal of the collimator cells, as compared to that along the side. This is not
accounted for as it is not possible to specify this angle during pointing.

 \begin{figure*} [th]
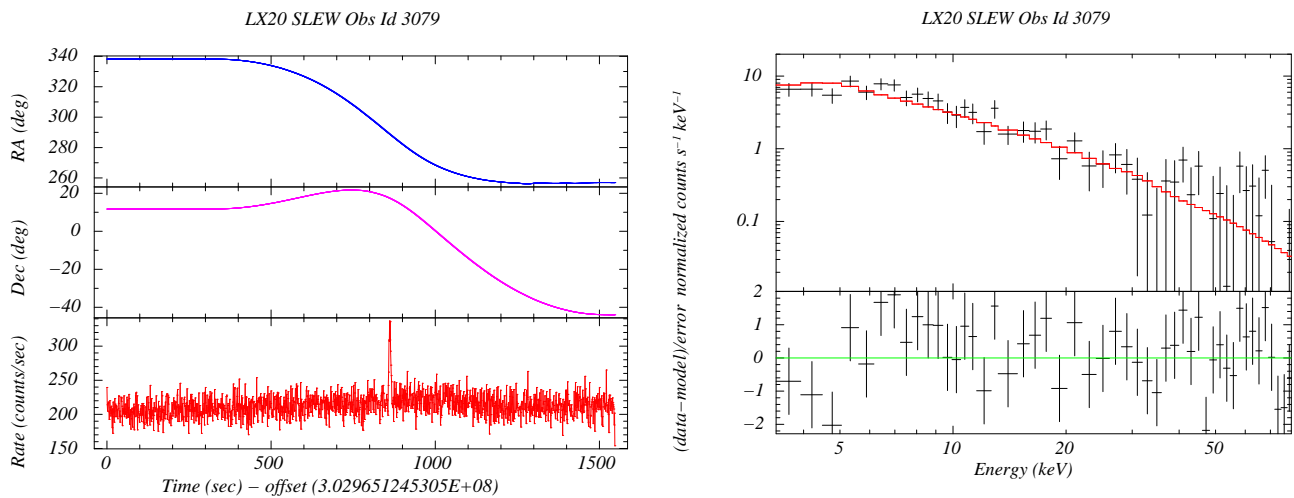

\centering
\includegraphics[height=1.01\columnwidth,angle=-90.0]{fig26a.eps}
\includegraphics[height=1.01\columnwidth,angle=-90.0]{fig26b.eps}
\caption{The light curve during a slew observation on August 8, 2019 is shown in
the left panel. The right panel shows the fit to the spectrum of the possible transient
source taken over the duration of the peak in the light curve.}
\label{fig:slew3}
\end{figure*}

\subsection{Possible Detection of a New Transient}

LAXPC can be used to scan the sky for new sources. Although, the pre-planned scan has not been attempted
so far, between two observations the satellite slews from one source to another and this
is similar to a scan. All slew observations have been analysed and during one of these
observations on August 8, 2019\footnote{ID: 20190808\_SLEW\_01234\_9000003079} a peak was seen in
the light curve around 13:13 UTC when the instrument was pointing at RA 288.717 and Dec 17.367. Figure~\ref{fig:slew3}
shows the light curve for the observation and the source spectrum taken during the peak.
The  spectrum was fitted to a powerlaw form with a systematics of 1\%.
The resulting fit (Figure~\ref{fig:slew3}) with a reduced $\chi^2$ of 1.2 yielded the index of $1.67\pm0.09$.
A ToO to observe this region was proposed and the observation was carried out on August 15, 2019,
but no signal was seen during that observation.

\section{Analysis Software}

Three different software are available to process the level-1 data to obtain science products.
These are LAXPClevel2DataPipeline (Section 5.1) written in C++ and LaxpcSoft written in Fortran.
The latter includes two related software, the basic software includes two Fortran programs,
{\tt laxpcl1} and {\tt backshift}
developed by TIFR team (Section 5.2) and  a suite of Fortran programs based on the basic programs for different
tasks developed by the AstroSat Science Support Cell at IUCAA (Section 5.3).
These are described in the following subsections.

\subsection{LAXPClevel2DataPipeline}

The LAXPC Level-2 data pipeline version 3.1\footnote{https://www.tifr.res.in/\~{}astrosat\_laxpc/LAXPC\_lvl2\_pipeline.html} was developed 
with support from ISRO Space Applications Centre (SAC),  Ahmedabad. The pipeline works
on the level-1 data downloaded from the Indian Space Science Data Centre (ISSDC) archive.
The pipeline generates various level-2 output data files for the scientific
studies and other ancillary files which are inherited from the level-1 data in FITS format. Before
producing these files, the pipeline checks for the data frame sequence order,
duplicate frames or possible frame corruption during data transfer. If needed, the data frames are
reordered.
The level-2 files generated for the scientific studies are :  event file, Good Time Interval (GTI) file,
lightcurve file and spectrum file. The other ancillary files inherited from the level-1 are : Time
Calibration Table (TCT) file, Make Filter (MKF) file giving information about the satellite
position etc., Low Bitrate Telemetry (LBT) house-keeping
file, orbit file and attitude file. Additionally, the pipeline also has independent routines to reprocess the
lightcurve and spectrum which can be generated as per requirements.
Currently, this pipeline has only limited support.

\subsection{{\tt laxpcl1} and {\tt backshift}}

These are two standalone Fortran programs\footnote{https://www.tifr.res.in/\~{}astrosat\_laxpc/LaxpcSoft.html} to process the level-1 data that are available
from the AstroSat archives. The readme files with the package give the detailed instructions
for usage. The tar file also includes the background files required for all background observations.
While the detector response files for all detectors are available from the LAXPC
website.
The level-1 inputs required includes the Event Analysis (EA) mode
files which give the record of each event that is detected and Broad Band counting (BB) mode files
giving the counts of various categories of events in a predefined time-bin. Apart from these, the .mkf file giving
the orbital and pointing parameters and the time calibration table (tct) file giving the conversion from
instrument time to UTC are also required. These programs can handle data from multiple orbits
after removing the overlapping part of data between consecutive orbits, but treat each detector
independently. This program also corrects for problems with frame sequence ordering as well
as event time ordering. The program also gives a statistics of the data processed, including
the total exposure, fraction of gaps in data and frame-loss that has been detected. In
most observations the frame-loss is less than 0.1\%, but in some cases it can be a few
percent or more. In such cases it may be necessary to discard the time intervals with
large frame-loss. Orbit-wise statistics of frame-loss is available in `.frame' file, which
can be listed using `grep Frame lxp2level2.frame'.

The program {\tt laxpcl1} is the main routine which generates the light curve,
spectrum, event file and GTI file giving the list of GTIs. All output files are in both ASCII and FITS format.
It also generates the background spectrum and light curve using the background model to estimate
the background contribution. A FITS file giving orbital parameters as required by the
tool {\tt as1bary}\footnote{http://astrosat-ssc.iucaa.in/?q=data\_and\_analysis} to apply
the barycentric corrections to time is also generated.
The channel range and anodes to be used can be specified.
To suppress some oscillations in spectrum, two channels are binned together for LAXPC10 and LAXPC30, while
for LAXPC20, four channels are binned. Thus the original spectrum in 1024 channels is reduced to
512 or 256 channels. However, this binning is done at the end and the input channel range to
be specified to {\tt laxpcl1} is in the full range of 0--1023. This should be accounted for while
specifying the input.
The events in event file are of two types,
single events, where all energy is deposited in one anode, and double events, where the energy
is deposited in two anodes, of which at least, one energy is in Xe K X-ray range. There is an
option to exclude double events. This may be useful in some cases as the response is better
defined for these, though the efficiency may be reduced to some extent (Antia et al.~2017).
Since the ULD threshold is applied to each anode separately, the double events can have
energies exceeding ULD and in principle, it is possible to study the spectrum up to about
110 keV, and these are included in the response files. However, the efficiency of detector
is very low for these events.  Since the event files
can be large, there is an option to suppress writing these files. This program generally
requires two passes, one to generate the GTI list and another to do the calculations using
this GTI file, which could be edited to choose any subset of intervals. To generate the GTI file
it is advised to use a time-bin of at least 1 s. The final light curve can be generated for
any time-bin, though there is some limit on dimension which may truncate the light curve when
very small time-bin is used. If the GTI file is already available then only one pass through
{\tt laxpcl1} is needed.

The program {\tt backshift} is used to account for gain shift between the source and background
observation after running {\tt laxpcl1}. It uses the log of gain to estimate the shift in gain and corrects the
background spectrum to match the gain during source observation. The correction is applied
both to the background spectrum and light curve. The background observation to be
used has to be specified in an input file for {\tt laxpcl1}. In general, the background observation
closest in time to source observation is to be used and {\tt backshift} gives recommendations
about which background file can be used. It also gives recommendation about which response
file has to be used to fit the spectrum. There are four sets of response files, the default is
for all events and anodes. This may be the name that is recommended by {\tt backshift}. Other set
of responses are for single event mode with `SE' in the name. The remaining two sets are
for only the top layer, `L1' and `L1SE' for all events and only single events, respectively.
There is also an option to remove diurnal variation in the light curve, which would not affect
the spectrum. For LAXPC30, response files are available for different densities and the program
also recommends which response needs to be used. In some cases, neighbouring density may
also be tried. For LAXPC20, the program also recommends the value of offset that may
be used in the {\tt gain fit} while fitting the spectrum. The offset may be frozen to this value
(and slope to 1.0) if there
is a difficulty in estimating the gain parameters using {\tt gain fit}.

For faint sources or sources with soft spectrum, it may be advisable to use the single event mode
with energy restricted to 3--20 keV. For this the option $\mbox{\tt ian}=1$ (for top layer),
$\mbox{\tt nul}=-2$ (to apply background fit for top layer)
and $\mbox{\tt iev}=1$ or $-1$ (for selecting single events only) should be used.
The energy range to be used can depend on the source.

\subsection{Software with Individual Routines}

A wrap around software to the primary one described above
has been developed for ease of certain kind of analysis
of LAXPC data. The software can be obtained from
the AstroSat Science Support Cell web-page
\footnote{http://astrosat-ssc.iucaa.in/?q=laxpcData}.
Details of the software and instructions for use
are given in  README files included in the software.
Here we outline some of the basic highlights and
the functionality of the software.

The software can produce a combined merged clean
event file (typically called level2.event.fits)
for all three LAXPC units which is
time sorted, allowing for ease of timing analysis.
Apart from Time, the other columns of the event file
are Layer number, LAXPC unit, Channel and Energy. The
Energy of the event is estimated using the
channel to energy conversion from the appropriate response file.
This allows the user to make appropriate selections, either
using the {\tt ftools} command {\tt fselect}, or by using the other routines
of the software. The Event file also contains as separate
data unit, the appropriate names of the response files for
the three LAXPC units as a function of time for the observation.

A number of individual routines are provided which can run
from the command line with flags as inputs. Hence they allow
for easy customized scripting by the user. The spectra and lightcurves
generated are compatible with high level {\tt HEASoft} tools such
as {\tt powspec, lcurve, Xspec, crosscor} etc. The
individual routines can be used to obtain:

\begin{enumerate}

\item A Good Time Interval (GTI) file that takes
into account earth occult and SAA passage.

\item A merged orbit file that is required to
make Barycentric corrections.

\item Lightcurves at a given time-bin for
user specified multiple energy bands with a GTI file as a input.
Lightcurves can be generated for different combination of the
LAXPC units and for all layers or only for the first layer.

\item Spectra for each of the three units for the input
GTI file. Spectra can be generated for all layers or only
for the first layer. Appropriate response files are copied
from the software database to the working directory and
the `RESPFILE' keyword in the spectra files is updated.

\item Estimated background spectra for each LAXPC
for the input GTI file. The background estimation can be
done either based on blank sky observations closest to data
or gain. All or first layer can be chosen.
The `BACKFILE' keyword in the spectra file is
changed to the resultant background file.

\item Estimated background lightcurve corresponding
to the extracted lightcurve.

\item GTI file for providing the time intervals
when the flux given in an input lightcurve is within some user
specified values. This is useful to generate flux resolved spectroscopy.

\item Time-lag, fractional r.m.s, coherence and
intrinsic coherence as a function of both frequency and energy.
These are computed  directly from the event file rather than
lightcurves and hence is computationally efficient when performing
high frequency analysis. The routine also provides the
power spectrum along with the
dead-time corrected Poisson level. A subsidiary routine rebins
the power spectrum and converts it into {\tt Xspec} readable format,
which allows the user to fit the power spectrum using {\tt Xspec}
models and flexibility.

\item Dynamic Power spectra which are power spectra
computed for consecutive time intervals. This is useful to
see the rapid variation in Quasi-periodic oscillation (QPO) frequency.

\item Estimate of background spectra and lightcurve
using an alternate method applicable for faint sources as described
by Misra et al.~(2020).

\end{enumerate}

\section{Science Goals of LAXPC Instrument: Status of their Realisation so Far}

AstroSat was conceived as a multiwavelength observatory for 
simultaneous observations of galactic and extragalactic sources in Visible, UV and X-ray 
bands using a suite of four co-aligned instruments for studies of different classes of 
galactic and extragalactic objects.  Three co-aligned X-ray instruments, sensitive in the 
X-ray energy region 0.5--150 keV were designed to investigate temporal and spectral 
characteristics of the sources to elucidate their nature and probe the complex radiation 
process operating in them.

The LAXPC instrument was designed to measure the intensity variations of bright X-ray
sources over a wide time scale, from 0.1 ms to minutes, days and months. The X-ray
binaries, with a neutron star or a black hole as the compact object, were special targets for the
LAXPC studies, as they exhibit a range of periodic and aperiodic variations on almost all time scales.
Accurate determination of continuum energy spectra of X-ray binaries to decipher the dominant
radiation processes as well as to search for the presence of weak spectral features known as Cyclotron lines
that provide a measurement of the magnetic field of the neutron star, is another major objective of
LAXPC.
To achieve these objectives, LAXPC tags the arrival time of every detected photon with an accuracy of 10 $\mu$s
and also achieves a moderate energy resolution over the entire range of 3--80 keV.

In this section we demonstrate how the various science goals of LAXPC instrument have been met by giving some
representative examples in each case. More detailed discussion of scientific result
is covered by Yadav et al.~(2020).
The LAXPC instrument was designed with the following objectives (Agrawal et al.~2017):

\begin{enumerate}
\item \textbf{Detailed studies of stellar-mass and supermassive black holes:} Several black hole
sources in both mass ranges have been studied. For example,
using the combined SXT and LAXPC spectrum, Mudambi et al.~(2020) have
derived the spin of the black hole in LMC X-1.
Similarly, Pahari et al.~(2018) and Sridhar et al.~(2019) have estimated the mass and spin of black holes
in 4U 1630--472 and MAXI J1535--571.
An example of a study of supermassive black hole is the blazar
RGB J0710+591 by Goswami et al.~(2020) using the multiwavelength capability of AstroSat
by combining data from UVIT, SXT and LAXPC.

\item \textbf{Studies of periodic (pulsations, binary light curves) variabilities in X-ray sources:}
The LAXPC instrument with its high time resolution 
capability is ideally suited to measure the spin periods of the X-ray pulsars to a 
high precision and deduce their derivatives. The spin-up or spin-down rates of these
pulsars is determined by many factors, including accretion and the magnetic field,
which in turn can be studied by the measured changes in the spin period.
Amin et al.~(2020) measured the 
spin period  and spin-up rate of the neutron star in the LMXB source 3A 1822--371.
The shape of the emitted pulse depends on modes of accretion, geometry of
accretion column and configuration of its magnetic field with respect to an observer’s line of sight.
Therefore, such studies offer insight into the physical processes in the vicinity of the pulsar.
For example, Mukerjee et al.~(2020a) studied the spectral and timing properties of GRO J2058+42
using data from AstroSat during a rare outburst in April 2019.
LAXPC data have been used to study pulsation over a wide range of period from 2.3 ms for
a transient accretion powered millisecond X-ray pulsar SAX J1748.9--2021 (Sharma et al.~2020) to 604 s for 4U 1909+07 (Jaiswal et al.~2020).
LAXPC is well suited for studying the pulsar period evolution and hopefully
more such studies will emerge in the future.
LAXPC also offers the possibility of inferring the evolution of the 
binary period by studying the binary light curve over a long base line of years. Using 
LAXPC observations of Cyg X-3 covering nearly one year, Pahari et al.~(2018) determined the 
binary orbital period to be $17253.56 \pm 0.19$ s.

\item \textbf{Studies of QPOs  and aperiodic variabilities in X-ray sources:}
LAXPC data have been used to study QPOs in a number of X-ray sources covering a wide
range of frequencies. For example, Belloni et al.~(2019) and Sreehari et al.~(2020) 
detected a clear HFQPO in well known black hole binary GRS 1915+105, whose frequency varied 
between 67.4 and 72.3 Hz.
A QPO  of 90 mHz  centroid frequency was detected for the first time by LAXPC in
GRO J2058+42 during its rare outburst of 2019 (Mukerjee et al. 2020a).
Apart from QPOs, the LAXPC instrument with its high time-resolution also allows the study
of the time lag between different energies of X-rays. Because of wide energy coverage these
studies can be extended to energies of 30 keV and above.
Misra et al.~(2017) have analysed LAXPC data from 
Cyg X-1 in the hard state to derive time lag between Soft (5--10 keV) and 
Hard (20--30 keV) photons which increase with energy for both the low and high frequency 
components. The event mode LAXPC data allowed them to perform flux resolved spectral 
analysis on a time-scale of 1~s, which clearly shows that the photon index increased 
from 1.72 to 1.80 as the flux increased by nearly a factor of two.
Apart from QPO, the neutron star X-ray binaries show thermonuclear X-ray bursts which
have also been studied by LAXPC. 
Verdhan Chauhan et al.~(2017) used the LAXPC observation of the LMXB 4U 1728-24 on 
March 8, 2016 to study a typical Type-1 burst of about
20 sec duration. The dynamical power spectrum of the data in the 3--20 keV band, shows the presence of a
burst oscillation whose frequency increased from 361.5 to 363.5 Hz.
These
results demonstrate the capability of LAXPC instrument for detecting millisecond variability
even from short observations.
Similarly, Devasia et al.~(2021) have studied thermonuclear bursts
in Cyg X-2. They have carried out energy resolved burst profile analysis as well as
time resolved spectral analysis for each of the 5 bursts that were observed.

\item \textbf{Low to moderate spectral resolution studies of continuum X-ray emission and CRSF:}
One of the prime objectives of LAXPC is to measure the continuum
energy spectra of Neutron star binaries to a high precision for detecting usually faint
signal of the Cyclotron lines. Using the LAXPC observations, Cyclotron lines have been
discovered in at least half a dozen pulsars. In section 4.3 detection of CRSF
in the spectra of several pulsars observed with the LAXPC, has been discussed
demonstrating the LAXPC capability for such investigations.
For example, Mukerjee et al.~(2020a) carried out a
phase resolved study of CRSF in GRO J2058+42, to find significant variation in CRSF energy
with pulse phase. Bala et al.~(2020) studied secular variation in energy of CRSF in
Her X-1 by comparing the result with earlier observations..

\item \textbf{Search for transient X-ray sources by surveys in a limited region of the galactic plane:}
So far a systematic survey has not been carried out, but between two science observations
the satellite slews from one source to another. The slew data have been scanned for
potential sources, but most features were found to be associated with known sources, except
for the one instance described in Section 4.5.

\end{enumerate}

The main goals of LAXPC have been realised to a  certain extent. However, the
detection of new transient sources in limited  regions of the galactic plane through
a dedicated survey is not yet planned.
A large fraction of the LAXPC and SXT data still remains to be analysed.
This could be achieved
by encouraging and involving  more number  of students and researchers
in data analysis and improving our  understanding of calibrations issues
pertaining to instrument gain, background, spectral response etc.
The phenomenon of kHz QPOs, a major discovery by RXTE in about 15 LMXBs, has
not been revisited by AstroSat even though it is ideally suited for this
study. The multiwavelength study was a prime objective of AstroSat to understand the
relationship between radiation in different bands in the AGNs i.e., is the UV flux from
AGNs generated by the reprocessing of X-rays in the accretion disk? In X-ray
binaries one measures time lag of hard X-rays with respect to the soft photons to infer
if hard X-rays are reflection component due to up scattering of soft photons on the
surface of the hot disk. It is hoped that some of these gaps in the AstroSat/LAXPC science
results will be addressed in the coming years.

\section{Summary}

AstroSat has completed five years of operation, during which more than 2000 different
pointings and over 1000 distinct sources have been observed. The data for most
observations are now available from the AstroSat archive. By now only one of the
three detectors, i.e., LAXPC20 is working nominally. The response of detectors is
reasonably understood and is stable. The detector background has been increasing
with time and some diurnal variations have been seen both in the background and in fitted
parameters to source spectrum. The variation in background puts a limit on the source
flux of about 1 mCrab, below which it would be difficult to study the source.
Even at this limiting flux it is possible to fit the spectrum to a reasonable extent
and to study pulsations. LAXPC has successfully detected CRSF in several sources. The
detection of these features can be confirmed by looking at the ratio of source to the Crab
spectrum.

To account for drift in the gain of detectors, it is recommended
to use {\tt gain fit} in {\tt Xspec} while fitting the spectrum, even when the recommended
response is used.
The fitted slope and offset should be compared with the values shown in Section 2.3
to check that the correction is in a reasonable range. Significantly different values
would imply that {\tt gain fit} has fitted some other spectral variation. In such cases
the slope should be fixed at 1 and the offset should be fixed at the value recommended
by {\tt backshiftv3}.
A few observations may have a significant
frame-loss during data transmission and it may be necessary to discard the data during these
time intervals. For faint sources it is advisable to use only the top layer of the detector
to reduce the background. Because of a relatively large field of view, there is a possibility
of another source in the field of view and this should be checked before analysing the
LAXPC data.

Many X-ray sources have been studied using LAXPC data resulting in more than 40 publications.
The scientific results are
summarised in a companion paper (Yadav et al.~2020).

\section*{Acknowledgment}
We acknowledge the strong support from Indian Space Research Organization (ISRO) in various
aspects of instrument building, testing, software development and mission operation
and data dissemination.  We acknowledge support of the scientific and technical staff of the LAXPC instrument team as well as staff of the TIFR Workshop in the development and testing of the LAXPC instrument.


\end{document}